\documentclass[letterpaper,12pt]{article}

\pdfoutput=1 
\usepackage{jheppub} 
\usepackage{subcaption}

\newcommand{\phir}{f}
\newcommand{\phia}{\tilde{\phi}}
\newcommand{\Erdot}{\dot{\mathbf{E}}_\text{r}}
\newcommand{\Brdot}{\dot{\mathbf{B}}_\text{r}}

\newcommand\beq{\begin{eqnarray}}
\newcommand\eeq{\end{eqnarray}}

\newcommand{\mybar}[1]%
        {\kern 0.6pt\overline{\kern -0.6pt#1\kern -0.6pt}\kern 0.6pt}
\DeclareMathOperator{\sech}{sech}
\newcommand{\norm}[1]{\lvert #1 \rvert}
\newcommand{\B}{\mathbf{B}}
\newcommand{\E}{\mathbf{E}}

\newcommand{\Br}{\mathbf{B}_\text{r}}
\newcommand{\Er}{\mathbf{E}_\text{r}}
\newcommand{\J}{\mathbf{J}}

\newcommand{\x}{\mathbf{x}}

\newcommand{\pt}{\partial_t}

\newcommand{\dd}{\text{d}}

\newcommand{\op}{\omega_\text{p}}

\newcommand{\A}{\mathbf{A}}

\newcommand{\rz}{\hat{\mathbf{x}}_z}

\newcommand{\g}{\left(\frac{C\beta}{\pi f_a}\right)}
\newcommand{\gp}{\left(\frac{C\beta\phi}{\pi f_a}\right)}
\newcommand{\Bs}{\mathbf{B}_\text{s}}
\newcommand{\Es}{\mathbf{E}_\text{s}}

\hypersetup{
    colorlinks=true,   
    linkcolor=red,     
    citecolor=blue,    
    filecolor=magenta, 
    urlcolor=blue      
}

\makeatletter
\def\l@subsubsection#1#2{}
\makeatother    

\advance\footnotesep 1.5pt

\begin{document}
\title{Energy conservation and axion back-reaction in a magnetic field}
\author[1]{Srimoyee Sen,}
\emailAdd{srimoyee08@gmail.com}
\author[1]{Lars Sivertsen,}
\emailAdd{lars@iastate.edu}
\affiliation[1]{Department of Physics and Astronomy,  Iowa State University, Ames IA 50011}
\abstract{Axion clumps  in an external magnetic field can emit electromagnetic radiation which causes them to decay. In the presence of a plasma, such radiation can become resonant if the clump frequency matches the plasma frequency. Typically, the decay or back-reaction of the clump is ignored in the literature when analyzing such radiation.
In this paper we present a self consistent, semi-analytic approach which captures axion back-reaction using energy conservation. 
We find that inclusion of back-reaction changes the clump frequency over time enabling clumps with a range of different initial frequencies to become resonant at some point in their time evolution.
}
\maketitle
\section{Introduction}
Axions are pseudo-scalar particles originally introduced to solve the strong CP problem in QCD \cite{PhysRevLett.38.1440, PhysRevLett.40.223, DINE1983137}. In many models, the axion (and axion-like particles (ALP)) interact weakly with standard model particles and are therefore considered to be suitable candidates for dark matter \cite{Preskill:1982cy, Abbott:1982af, Kim:1986ax, Cheng:1987gp, Raffelt:1990yz, Duffy:2009ig,Berezhiani1991,1992SvJNP..55.1063B,1994PAN....57..485S,1996PAN....59.1005S,KHLOPOV1999105,PhysRevD.99.064049,PhysRevD.99.104070,Odintsov_2020,Oikonomou_2022}. 
Due to their bosonic statistics, axions can form coherently oscillating Bose-Einstein condensates (BEC). These BECs are a consequence of axions clumping in space owing to their gravity and or their self interaction \cite{Visinelli:2017ooc,Zhang:2018slz,Guth_2015}. The clumps are well described by a localized spatial profile for the axion field, which can be obtained by solving the classical equations of motion for it \cite{Zhang:2018slz,Visinelli:2017ooc,Eby:2015hyx,Zhang:2020bec,Zhang_2021}. Since axion number is not conserved, the clumps can decay via scalar radiation. This causes the clumps to have a finite lifetime. However, this decay can be significantly suppressed for certain axion potentials \cite{Eby:2015hyx,PhysRevLett.117.121801,Visinelli:2017ooc} giving rise to long-lived clumps.

Axions can also couple to electromagnetic fields and send out electromagnetic radiation through several different processes which in turn can cause the clumps to decay. For example, an axion particle can decay to two photons which allows axion clumps to radiate via spontaneous or stimulated emission. For ultralight axion clumps however, the former is highly suppressed  \cite{Sen:2021mhf} whereas the latter isn't. The decay time for stimulated emission can be small, for example of the order of a few seconds for ultralight axions of mass $10^{-11}$ eV. Similarly, in the presence of a background electromagnetic field, an axion particle can convert to a photon. This causes an axion clump in a electromagnetic (EM) field to emit EM radiation. To see how this comes about, one has to solve the classical equations of motion for the axion and the electromagnetic fields simultaneously. Maxwell's equations in this case get augmented with an oscillating current source term originating from the axion clump, which acts like an antenna contributing to EM radiation. 

All of these processes take away energy from the axion clump depleting the clump. This in turn reduces the EM radiation coming out of these clumps. Typically when considering electromagnetic radiation in these problems, one considers the axion amplitude to be independent of time ignoring the back-reaction of the radiation on the axion clump itself. 
This is justified when axion-photon coupling is weak so long as one is only interested in the leading order EM radiation coming out of the clump.
One can estimate the axion decay timescale by taking a ratio of the total initial energy stored in the axion clump to the total radiated power. For example, in \cite{Arvanitaki_2010} it is mentioned that for an axion photon coupling of $\frac{C\beta}{\pi f_a}$ where $C$ is an order $1$ number and $f_a$ is the axion decay constant, the decay time of an axion clump with axion mass $m_a$ in the presence of a uniform background magnetic field $B$ is given by
\beq
\tau\sim\left(\frac{\pi f_a}{C \beta}\right)^2 \frac{m_a}{3B^2}.
\label{tau}
\eeq
This procedure may give approximately correct estimates for the decay time for the axion clumps. However, there are several interesting and phenomenologically exciting features of axion radiation which cannot be captured in this method. For example, consider the findings of \cite{Amin2021,Sen:2021mhf} where it was shown that the EM radiation efficiency of an axion clump in external EM field depends on a combination of the clump frequency and its spatial extent. Clumps that were too large or too small compared to the wavelength of the radiation would radiate minimally, whereas there would be efficient (resonant) radiation when the two values were close. Strictly speaking, this conclusion is accurate only when the decay of the axion clump (axion back-reaction) is ignored. As observed in \cite{Zhang:2020bec,Visinelli:2017ooc} the decay of axion clumps causes their frequency, their spatial extent and the wavelength of EM radiation to change over time. This makes it possible to imagine scenarios where a poorly radiating clump at a certain instant in time can become resonant at a later instant by altering its frequency and spatial extent as it radiates. Similarly, a clump that is radiating resonantly at a certain instant in time can move out of resonance as time passes. This dynamical change in resonance condition cannot be captured in an analysis which ignores back-reaction where the axion amplitude and its frequency are assumed to be constant in time. Our goal in this paper is to address this by taking into account axion back-reaction. Our framework is specifically designed to address EM radiation from clumps in a background magnetic field. It should be possible to extend this approach to stimulated emission, which however is beyond the scope of this paper.


In the process of constructing our framework we verify the estimate given by Eq. \ref{tau} while also capturing how clumps can move in and out of resonance with time as they radiate. Note that, some previous analysis, e.g. \cite{Amin2021} has taken into account the effect of axion back-reaction for sufficiently large axion-photon coupling where stimulated emission dominates. Their analysis involves complete numerical simulation of axion-photon equations of motion as necessitated by a strong axion-photon coupling. 
Such numerical simulation is unnecessary when the coupling is weak as is the case in this paper.
Instead, here we develop a semi-analytic perturbative approach to taking into account axion back-reaction based on energy conservation. This procedure captures axion clump decay and the resulting decay of electromagnetic radiation in the limit of weak axion-photon coupling while avoiding the cost of a full axion-photon numerical simulation. 


As stated before, axion clumps can also radiate by emitting scalar(axion) waves. The stability of relativistic axion clumps against scalar radiation was analyzed in \cite{Zhang:2020bec} and it was found that for certain types of axion self interactions the decay is suppressed sufficiently so as to produce long living axion clumps. Here the axion decay timescale is given by $\tau_{\text{scalar}}\gg m_a^{-1}$. For such clumps, it may be appropriate to neglect the effect of scalar radiation on the clump and to only consider the effect of electromagnetic radiation to describe the time evolution. This is the regime where we choose to work in, i.e. $\tau_{\text{scalar}}>\tau_{\text{EM}}>\tau_{\text{rad}}$ where $\tau_{\text{EM}}$ is the clump decay timescale due to EM radiation and $\tau_{\text{rad}}$ is the time period of radiated EM waves. As we will see, $\tau_{\text{rad}}\sim m_a^{-1}$.
In principle, a complete analysis of axion clump decay should include both the effect of the axion radiation and electromagnetic radiation. Since the goal of this paper is to describe the effect of electromagnetic radiation and the corresponding back-reaction, with a few exceptions we mostly restrict ourselves to parameter ranges where scalar radiation is suppressed compared to the EM radiation. 

The organization of this paper is as follows: In the next section we outline the perturbative semi-analytic approach to taking into account axion back-reaction. In the following two brief sections, section \ref{axion profile section} and section \ref{emag}, we review axion clump solutions and electromagnetic radiation from them. In the following section, section \ref{applications}, we highlight a few processes where the effect of backreaction can have dramatic effects on the time dependence of radiation, including resonant radiation. This is followed by results which demonstrate the same. We conclude with a discussion on the decay time-scale of axion clumps while outlining the regime of validity of our approach. 



\section{The axion-photon equations and backreaction}\label{argument}
In this section we will present the main ideas behind the semi-analytic approach we take to account for backreaction. Our calculations hold for weak axion-photon coupling in a regime where the axion decay timescale due to electromagnetic radiation is large compared to the time period of outgoing radiation. The axion-photon Lagrangian is given by
\begin{align}
\mathcal{L} = -\frac{1}{4}F^{\mu\nu}F_{\mu\nu}+J_m^\mu A_\mu +\frac{C\beta}{4\pi f_a}\phi\epsilon^{\mu\nu\lambda\rho}F_{\mu\nu}F_{\lambda\rho}+\frac{1}{2}(\partial_\mu\phi)(\partial^\mu\phi)-V(\phi)+....
\label{lag}
\end{align}
In the latter parts of this paper explain the regime of validity of our calculation in terms of the parameters of this Lagrangian.
In the Lagrangian above, $\phi(\x,t)$ is the pseudo scalar axion field with mass $m_a$, $F^{\mu\nu} = \partial_\mu A_\nu-\partial_\nu A_\mu$ is the electromagnetic field tensor with corresponding gauge field $A_\mu$, $V(\phi)$ is the effective axion potential, and $J_m^\mu$ is a background matter current, if any. Furthermore, $C$ is a model dependent parameter with $C\sim 1$, while $f_a$ is the axion decay constant, and $\beta$ is the electromagnetic fine structure constant. For the QCD axion one has $\beta\sim1/137$, while for axion-like particles $\beta$ can take other values. The axion mass for a particular potential can be obtained by $m_a=\frac{\partial^2V}{\partial\phi^2}\big|_{\phi=0}$.
If one wants to understand axion dynamics in the absence of electromagnetic interaction, one has to set $C=0$ in the above Lagrangian. As we know, a finite number of axion particles can clump together to form axion clumps. Such clump solutions have been analyzed by several papers \cite{Zhang:2018slz,Visinelli:2017ooc,Zhang:2020bec,Eby:2015hyx} in the absence of electromagnetic coupling.  These solitonic solutions can be found by writing down axion EOM from the above Lagrangian and solving it for a fixed axion number or total energy. These solutions will play a crucial role in our analysis of axion back-reaction. 

If one wants to analyze axion clumps in the presence of electromagnetic coupling, one needs to write both axion EOM and Maxwell's equations

\begin{align}
&\dot{\E}=\nabla\times \B+\g(\dot{\phi}\B+\nabla\phi\times\E)-\J_m,
\label{EOM1}
\\
&\dot{\B}=-\nabla\times \E,
\\
&\nabla\cdot\E=\g\nabla\phi\cdot\B+J_m^0 ,\label{EOM2}
\\
&\nabla\cdot \B=0,
\\
&\pt^2\phi-\nabla^2\phi+\partial_\phi V(\phi)=-\g \E\cdot\B.
\label{EOM}
\end{align}
and solve them simultaneously. 

As can be seen in the above equations Eq. \ref{EOM1} and \ref{EOM2}, an axion clump can source electromagnetic fields. The matter current $J_m^{\mu}$ in Eq. \ref{EOM1} and \ref{EOM2} captures any background plasma in which the axion clump can be submerged and also any current source which could give rise to some background electromagnetic field. Since this section is meant to illustrate how we treat back-reaction, we restrict ourselves to a regime where there is no background plasma. The arguments we present here can easily be translated to the case where we have a plasma background. 

In the absence of a plasma, the matter current is taken to be only sourcing a constant magnetic field $\B_s= B_0\rz$. Therefore, we can simply write $\nabla \times {\bf{B}}_s=\J_\text{m}$ and $J_m^0=0$. Consider now the axion EOM \ref{EOM}. If this equation gives rise to coherently oscillating axion clump under certain conditions,
one can substitute this solution in Eq. \ref{EOM1} to find the resulting electromagnetic field.  It is easy to see that an oscillating axion clump in the presence of a background magnetic field will act like an oscillating current source. This will result in propagating electromagnetic radiation which will take away energy from the axion clump. 


In similar spirit to \cite{Amin:2021tnq,Sen:2021mhf} we will separate the total electromagnetic field in an external part sourced by the matter current $\Es$, $\Bs$ with $\Es=0$, and that sourced by the axion which we denote as $\Er$ and $\Br$ such that $\B = \B_s +\Br$ and $\E = \Er$. We will work in a regime where $|\Br|/|\Bs| \ll 1$ which effectively translates to a $ \frac{C\beta\phi}{\pi f_a}\ll 1$. This is the regime of parameter space where we can treat the axion source term as a perturbation. The EOM can now be written as
\begin{align}
&\Erdot(\x,t)=\nabla\times \Br(\x,t)+\g\dot{\phi}\Bs,
\label{dot(E)}
\\
&\Brdot(\x,t)=-\nabla\times \Er(\x,t),\label{dot(B)}
\\
&\nabla\cdot\Er(\x,t)=\g\nabla\phi\cdot\Bs,\label{nabla(E)}
\\
&\nabla\cdot \Br(\x,t)=0,\label{nabla(B)}
\\
&\pt^2\phi(\x,t)-\nabla^2\phi(\x,t)+\partial_\phi V(\phi)=-\g \Er(\x,t)\cdot\Bs.\label{eq_of_motion}
\end{align}
Here we have dropped $\Br$ from the source terms since $\Br\ll\Bs$ in the limit of $\gp\ll 1$. In principle we can solve these simultaneous equations numerically. However, this can be computationally expensive and hence unnecessary in the weak coupling limit. Of course, for large $\gp$ such numerical analysis is in fact warranted and was performed in \cite{Amin2021}. In the limit of small $\gp$ we employ an alternative approach as outlined in the next few paragraphs. As explained later in this paper, we restrict ourselves to $\phi\sim f_a$.
Therefore, a small $\gp$ effectively translates to small $\beta$ in our analysis. 

If we ignore back-reaction, one can simply solve Eq. \ref{eq_of_motion} by turning off $\gp=0$. These solutions were discussed in detail in \cite{Zhang:2020bec} where sets of solutions were obtained for different particle numbers. These solutions for different particle numbers also have different total energy. We will not discuss how one obtains these solutions in this section and instead assume that such solutions exist.
Substituting these axion clump solutions in Maxwell's equations one can obtain expressions for the radiated electromagnetic fields far away from the boundary of the clumpy solutions. This tells us exactly how much energy is being radiated away from the axion clump at a certain instant in time. 


With this let us now concentrate on energy conservation. 
From Eq. \eqref{dot(E)} and \eqref{dot(B)} we can relate the Poynting vector of the radiated EM field to the axion
photon coupling term in the Maxwell's equation as
\beq
\left(\Er\cdot\Erdot+\Br\cdot\Brdot\right)+\nabla\cdot(\Er\times \Br)=\g\dot{\phi}\Er\cdot\Bs
\eeq
Similarly, starting with \eqref{eq_of_motion}, multiplying by $\dot{\phi}$ and integrating over time we obtain
\beq
\pt\left(\frac{1}{2}\dot{\phi}^2+\frac{1}{2}(\nabla\phi)^2+V(\phi)\right)=-\g\dot{\phi}\Er\cdot\Bs.
\eeq
We can therefore relate the electromagnetic field energy, the radiated power and the energy density of the axion clump through
\beq
\left(\Er\cdot\Erdot+\Br\cdot\Brdot\right)+\nabla\cdot(\Er\times\Br)=\pt\left(\frac{1}{2}\dot{\phi}^2+\frac{1}{2}(\nabla\phi)^2+V(\phi)\right).
\label{energy}
\eeq
Note that the electromagnetic field sourced by the axion clump $\Er$ and $\Br$ is an alternating electromagnetic field with its frequency set by the axion clump frequency $\omega$.
At this point we can average both sides of the above equation Eq. \ref{energy} over a time period of the axion clump oscillation $\frac{2\pi}{\omega}$. The first two terms included in the parenthesis on the LHS vanish upon averaging. This is strictly true when the oscillating electric and magnetic fields have a constant amplitude. Since we will take into account back-reaction the electric and magnetic field amplitudes will change with time. However, this rate of change is slow in the $\gp\ll1$ limit, i.e. the rate of change of the electric and magnetic fields themselves go as powers of $\gp$. The second term which is the divergence of the Poynting vector does not vanish upon averaging even for constant electric and magnetic field amplitude. Therefore, we expect the first two terms in the parenthesis on the LHS to be smaller than the Poynting vector term on the LHS. Another way of stating the same is that the first two terms in the parenthesis are expected to be smaller compared to the Poynting vector term by factors of $(\omega\tau_{\text{EM}})^{-1}$ where $\tau_{\text{EM}}\gg \omega^{-1}$ is the electromagnetic decay timescale. If we now integrate over a volume of space much larger than the size of the axion clump, we find that the rate of change of the total axion clump energy is equal to the energy being radiated away
\beq
\int \dd^2\x\cdot (\Er\times\Br)\approx \int \dd^3x \,\,\pt\left(\frac{1}{2}\dot{\phi}^2+\frac{1}{2}(\nabla\phi)^2+V(\phi)\right)
\label{E2}
\eeq
up to leading order in $\gp$.

This causes the clump to decay. Note that the strength of the electric and magnetic fields themselves go as $\gp$. Therefore, the rate of change of the axion clump energy goes as  $\gp^2$. This implies that the clump amplitude will change at a rate proportional to powers of $\gp$. 

Having established the fact that the entirety of the radiated electromagnetic energy is sourced by the axion clump, we now have to determine how the axion profile changes with time in order to satisfy energy conservation. For this we have to consider the axion EOM 
\beq
\pt^2\phi(\x,t)-\nabla^2\phi(\x,t)+\partial_\phi V(\phi)=-\g \Er(\x,t)\cdot\Bs.
\label{back}
\eeq
Solving this equation in the limit of $C\beta=0$ and a finite particle number can yield localized axion clumps with large lifetimes depending on the details of $V(\phi)$. Such solutions will conserve particle number approximately in the limit $\gp=0$. As we will discuss in the next section, for these solutions the axion field can be written as $\phi\approx\phia\cos(\omega t) \phir(r)$ \cite{Amin2021,Zhang:2020bec,Schiappacasse_2018,Visinelli:2017ooc} where $\omega$ is the oscillation frequency, $\phia$ the central amplitude and $f(r)$ is some spatial profile representing the clump. For a certain fixed particle number, $\phia$, $\omega$ and $\phir$ are related to each other. Such a configuration also has a fixed energy. If we consider a different total energy, it will correspond to a different particle number with different central amplitude, frequency and spatial profile. 

Let's now turn on $\gp$ and solve Eq. \ref{back}. As we turn on $\gp$, we know from energy conservation equation that the axion clump amplitude and its frequency will change with time which in turn will make $\partial_\phi V(\phi)$ change with time. However, at every instant in time the term $\partial_\phi V(\phi)$ in Eq. \ref{back} is much larger than the term on the RHS $\g \Er(\x,t)\cdot\Bs$ which goes as $\beta^2$. Therefore, at leading order in $\beta$, the RHS of Eq. \ref{back} can be neglected such that we can replace the clump ansatz with the following expression 
\beq
\phi\approx \tilde{\phi}_{\text{av}}\cos(\omega_{\text{av}}t)f_{\text{av}}(r)
\eeq
where the subscript ``av'' stands for average over a period of $2\pi/\omega$. In other words, at a particular instant in time, over a time interval of $2\pi/\omega$ it is reasonable to approximate the solution to axion EOM with 
a solution found in the limit of $C\beta=0$ for a total energy consistent with Eq. \ref{E2}. This implies that the time dependence of the axion solutions can be obtained simply by allowing its profile and frequency to be time dependent in such a way that $\tilde{\phi}_{\text{av}}$ ($\omega_{\text{av}}$, $f_{\text{av}}$) at every instant of time satisfies Eq. \ref{back} with $C\beta=0$, while the difference in the total energy of any two configurations separated in time $\Delta t \gg \frac{2\pi}{\omega(t)}$ is set by $\int_0^{\Delta t} \dd t\int \dd^2x \,\,(\Er\times\Bs)$. This approach can be improved systematically by taking into account higher order corrections in $C\beta$.



Having outlined our strategy for taking into account back-reaction, we will now briefly review localized axion clump solutions and how electromagnetic radiation is computed from these solutions \cite{Amin:2021tnq,Sen:2021mhf}. Both will be used in computing the backreaction as we will outline in section \ref{applications}.  

\section{The axion profile}\label{axion profile section}
In this section we explain how we solve the axion EOM in the limit of $C\beta=0$ following the analysis of \cite{Zhang:2020bec}. These solutions will be crucial in capturing the time dependence of the axion clump and that of the EM radiation when we take into account backreaction in subsection \ref{power radiated}.
As mentioned in \cite{Zhang:2020bec} the axion field is well described by a spherically symmetric, harmonically oscillating profile 
\begin{equation}
\phi(\x,t) = \tilde{\phi}f(r)\cos(\omega t),\label{ansatz}
\end{equation}
with $\omega\sim m_a$ and $r = |\x|$. With this we can find localized clump solutions for $\phir(r)$ and the frequency $\omega$ for a fixed $\phia$ 
\begin{equation}
-\omega^2\phia\phir(r)\cos(\omega t)-\phia\nabla^2\phir(r)\cos(\omega t)+\partial_\phi V(\phia\phir(r)\cos(\omega t)) = 0.\label{sourceless_eq}
\end{equation}  
Here, the word `localized' refers to solutions that decay faster than $r^{-1}$ far away from the center of the clump. Note that the ansatz \eqref{ansatz} is just an approximate solution to the axion EOM. As outlined in \cite{Zhang:2020bec}, the exact solution instead has an infinite tower of modes $\phi(r,t) = \sum_{n=1}^{\infty}\phia_n\phir_n(r)\cos(n\omega t+\alpha_n)$, where $\alpha_n$ is a phase, $\phia_n$ is the amplitude of the $n$-th mode and $\phir_n$ is the radial profile for each mode, constrained by $\phir_n(0)=1$. However, up to moderately large amplitudes $\phia$, the lowest frequency mode dominates over the others \cite{Zhang:2020bec}, justifying the single frequency approximation. In our ansatz \eqref{ansatz} we denote $\phia_1=\phia$, $f_1(r)=f(r)$ and set $\phia_n=0$ for $n>1$. In order to make progress, one multiplies eq. \eqref{sourceless_eq} by $\cos(\omega t)$ and integrate it over one period of the clump $T = 2\pi/\omega$, yielding the result
\begin{equation}
\omega^2\phia\phir(r)+\phia\nabla^2\phir(r) = \frac{\omega}{\pi}\int_{-\pi/\omega}^{\pi/\omega}\dd t'\,\partial_\phi V(\phia\phir(r)\cos(\omega t'))\cos(\omega t'). \label{shooting}
\end{equation}
Note that here we are assuming that the amplitude $\phia$ and the profile $\phir(r)$ are constant in time. Later we will let the axion clump deplete due to electromagnetic interactions, and in order to still be able to use equation \eqref{shooting} we will have to work in a regime of parameters where $\phia$, $f(r)$ does not change much over a timescale $\sim \omega^{-1}$, that is, $\dot{\phia}\ll\omega\phia$ etc. This is going to be true for most of the parameter range as long as $\gp\ll1$ or $C\beta \ll1$ for $\phi\sim f_a$.

There is a wide range of potentials $V(\phi)$ that support spherically symmetric solutions of the form \eqref{ansatz}. One of these choices is the QCD instanton potential \cite{PhysRevLett.38.1440}
\begin{equation}
V_{\text{cos}}(\phi) = m_a^2 f_a^2(1-\cos(\phi/f_a)),
\end{equation}
derived from the dilute instanton gas approximation. We will from now on refer to this as the cos-potential. Note that, the solutions to the EOM specified by the ansatz in Eq \ref{ansatz}  are in fact only meta stable when one includes scalar radiation in clump analysis. However, as long as the decay timescale via scalar radiation is much longer than the oscillation time of the clump(inverse clump frequency) the ansatz of Eq. \ref{ansatz} is justified. In the absence of gravity and external electromagnetic fields, multiple sources have found the solutions to the cos-potential to have a lifetime of $\tau\sim 10^{3}m_a^{-1}$ \cite{Zhang:2020bec,Piette_1998}, due to the emission of scalar waves. For our axion back-reaction analysis, we too work with the ansatz in Eq. \ref{ansatz}. This is justified as long as the decay timescale through scalar waves is much larger than decay timescale via EM radiation.

Another example of a potential that allows the formation of spherically symmetric axion clumps is the tanh-potential \cite{Zhang:2020bec}
\begin{equation}
V_\text{tanh}(\phi) = \frac{1}{2}m_a^2 f_a^2\tanh^2(\phi/f_a). 
\end{equation}
It was found in \cite{Zhang:2020bec} that solutions for such a potential have a lifetime that is significantly longer than that of the cos-potential, with $\tau \sim 10^{6}m_a^{-1}$. These are just two examples of potentials that can support spherically symmetric localized solutions to the axion equations of motion \eqref{sourceless_eq}. Properties and examples of other such potentials can for instance be found in \cite{Zhang:2020bec,Amin:2021tnq}. We will in this paper focus on the cos- and tanh-potentials, although our methods can easily be applied to any potential that supports spherically symmetric localized solutions to \eqref{sourceless_eq} that are long lived. 

\begin{figure}
	\includegraphics{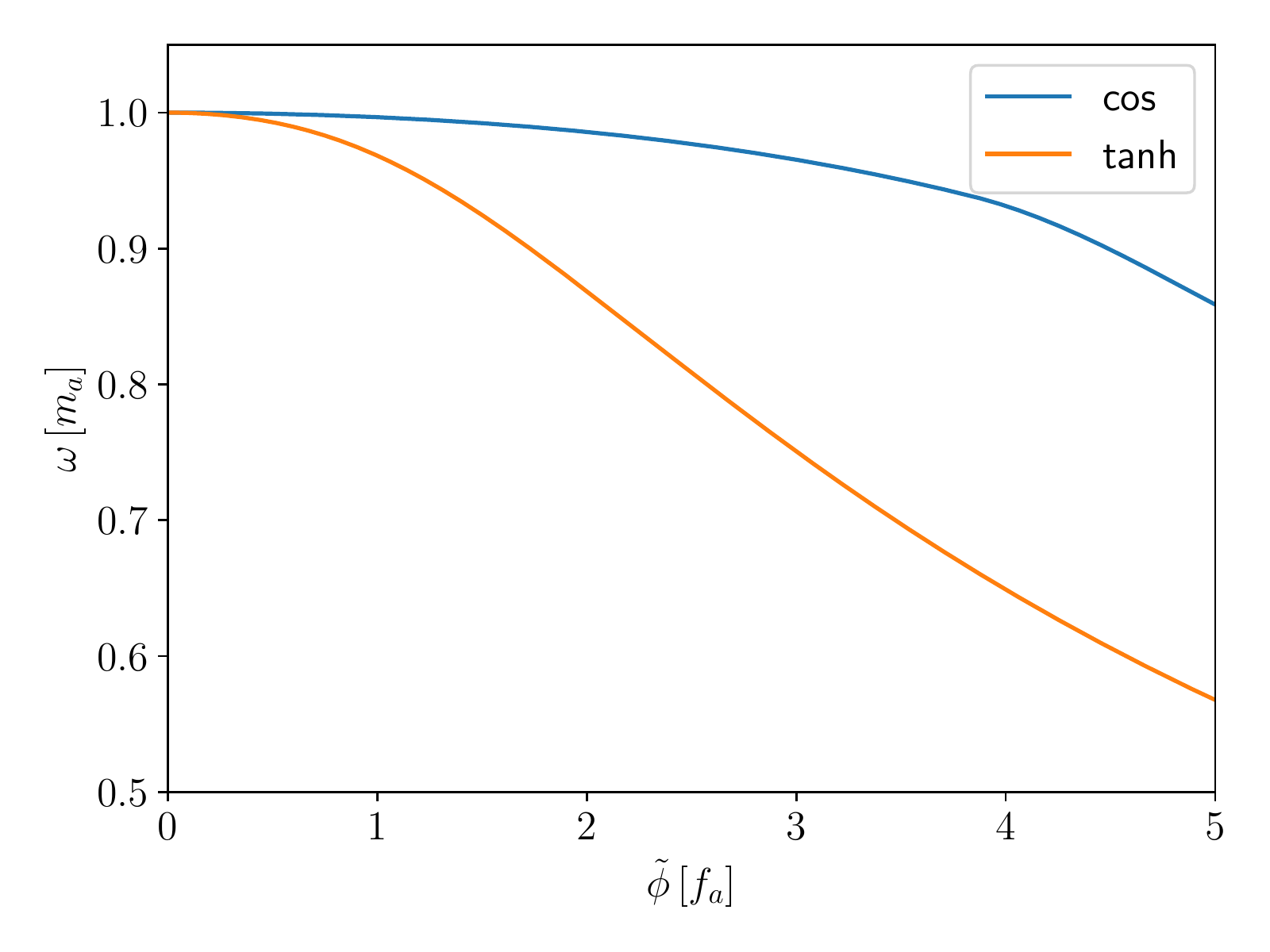}
	\caption{Clump frequency $\omega$ as a function of the central amplitude $\phia$ for the cos- and tanh-potential. }\label{omega_of_phi0}
\end{figure}

Once we have chosen the potentials we want to analyze, we solve equation \eqref{shooting} by what is often referred to as a shooting method \cite{alma990016848140102756}. That is, we pick one initial value for $\phia$, then we vary $\omega$ until we find $\phir(r)$ to be a monotonic function of $r$ that  vanishes faster than $r^{-1}$ far away from the clump. In finding the solutions we also impose the condition $\partial_r\phir(r)|_{r=0}=0$, since the derivative of $\phir(r)$ cannot be discontinuous at the center of the clump. Following this recipe, we obtain an explicit relationship between $\phia$ and $\omega$ shown in figure \ref{omega_of_phi0}. From the figure it is clear that the frequency $\omega$ increases as the central amplitude $\phia$ decreases, with $\omega \lesssim m_a$. This implies that whenever a clump depletes through radiation, its frequency will increase with time. 
There are so far two known branches of localized clump solutions of axion clumps that are stable towards small perturbations. One of these is the dense clump branch, characterized by having a size $R\sim m_a^{-1}$, frequency $\omega\lesssim m_a$ and central amplitude $\phia\sim f_a$ \cite{Zhang:2018slz,Visinelli:2017ooc}. For these solutions gravitational interactions can mostly be ignored, as we do here, since the self interactions are orders of magnitude larger \cite{PhysRevD.100.063002}. On the other hand, there is a set of dilute solutions with size $R\gg m_a^{-1}$ and central amplitude $\phia\ll f_a$ \cite{Grandcl_ment_2011,Eby:2015hyx,PhysRevLett.66.1659,Ure_a_L_pez_2002,Alcubierre_2003}. For these clumps gravity dominates over self interactions. Thus, having chosen to ignore gravity in our computation, we will not consider dilute clumps in this paper and restrict ourselves to $\phi\sim f_a$. As a result our weak coupling analysis is controlled by the smallness of $\beta$.

\begin{figure}
	\includegraphics{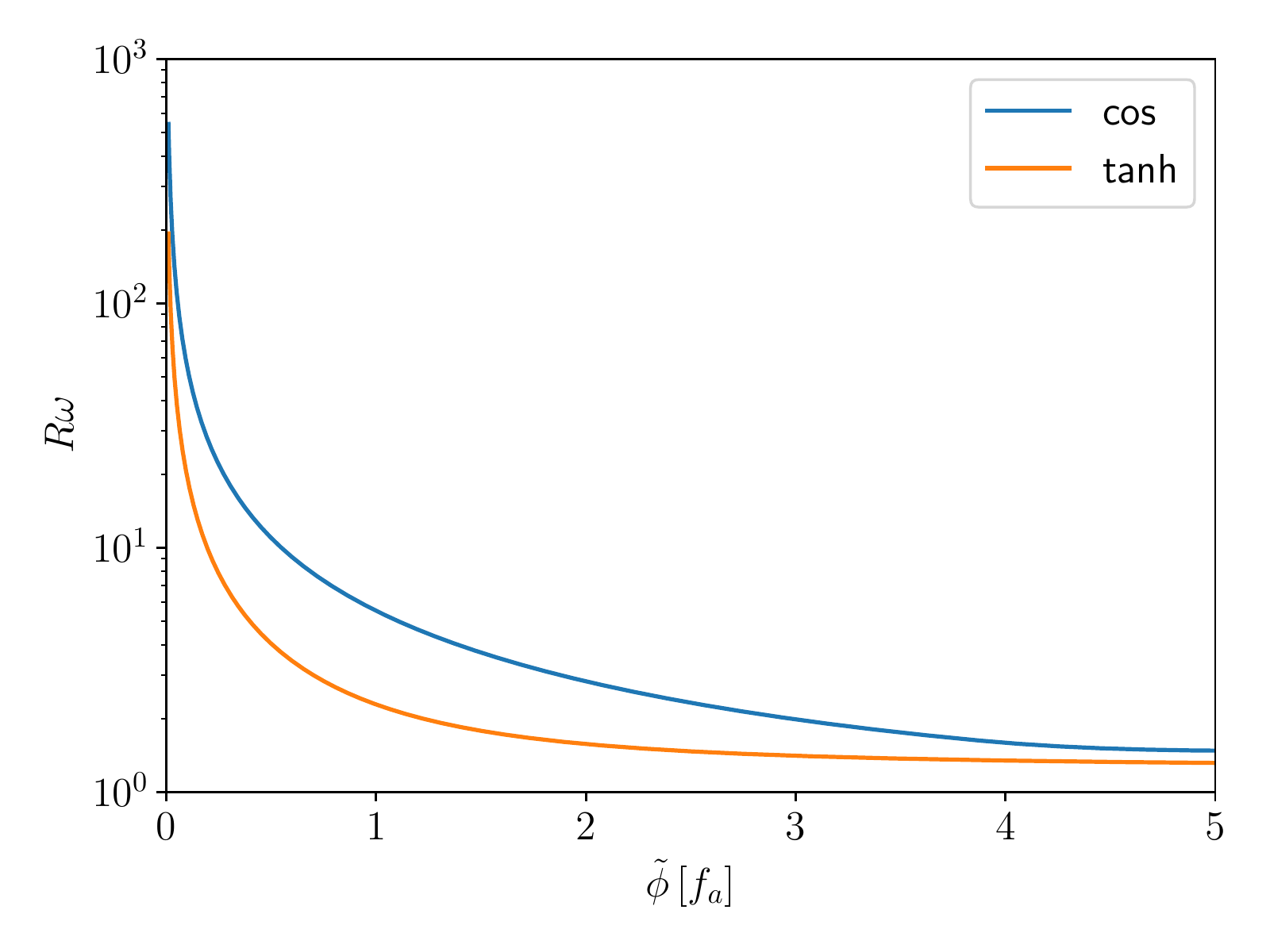}
	\caption{The product of the clump size $R$ (as defined in the text in Eq. \ref{ansatz}) and $\omega$ as a function of the central amplitude $\phia$ (in units of $f_a$) for the cos- and tanh-potentials.}\label{R_of_phi0}
\end{figure}

In \cite{Schiappacasse_2018} it was noted that these axion profile solutions $\phir(r)$ can be well approximated by a hyperbolic secant function $\sech(r/R)$ where $R$ is a free parameter relating to the clump size. This translates to the following ansatz for the axion clump
\begin{equation}
\phi(\x,t) = \phia\sech(r/R)\cos(\omega t)\label{ansatz_2},
\end{equation} 
and we can in our work determine $R$ by performing a curve fit to the obtained axion profile $\phir(r)$. The results for the clump sizes with central amplitudes in the range $0.01 f_a\leq\phia\leq 5f_a$ are shown in figure \ref{R_of_phi0}. Note that the size of the clump for $\phi\sim f_a$ for both potentials is given by $R\sim \omega_a^{-1}$, while the size increases by many orders of magnitude as $\phia\ll f_a$. In fact, for $\phia\lesssim f_a$, we have an inverse relationship between $\phia$ and $R$, with $R\sim \phia^{-1}$. We will avoid discussing very large $R$ regime since the corresponding small amplitude for the clump places it outside of the validity of our analysis

\subsection{Long wavelength instability}\label{Long}
\begin{figure}
	\includegraphics{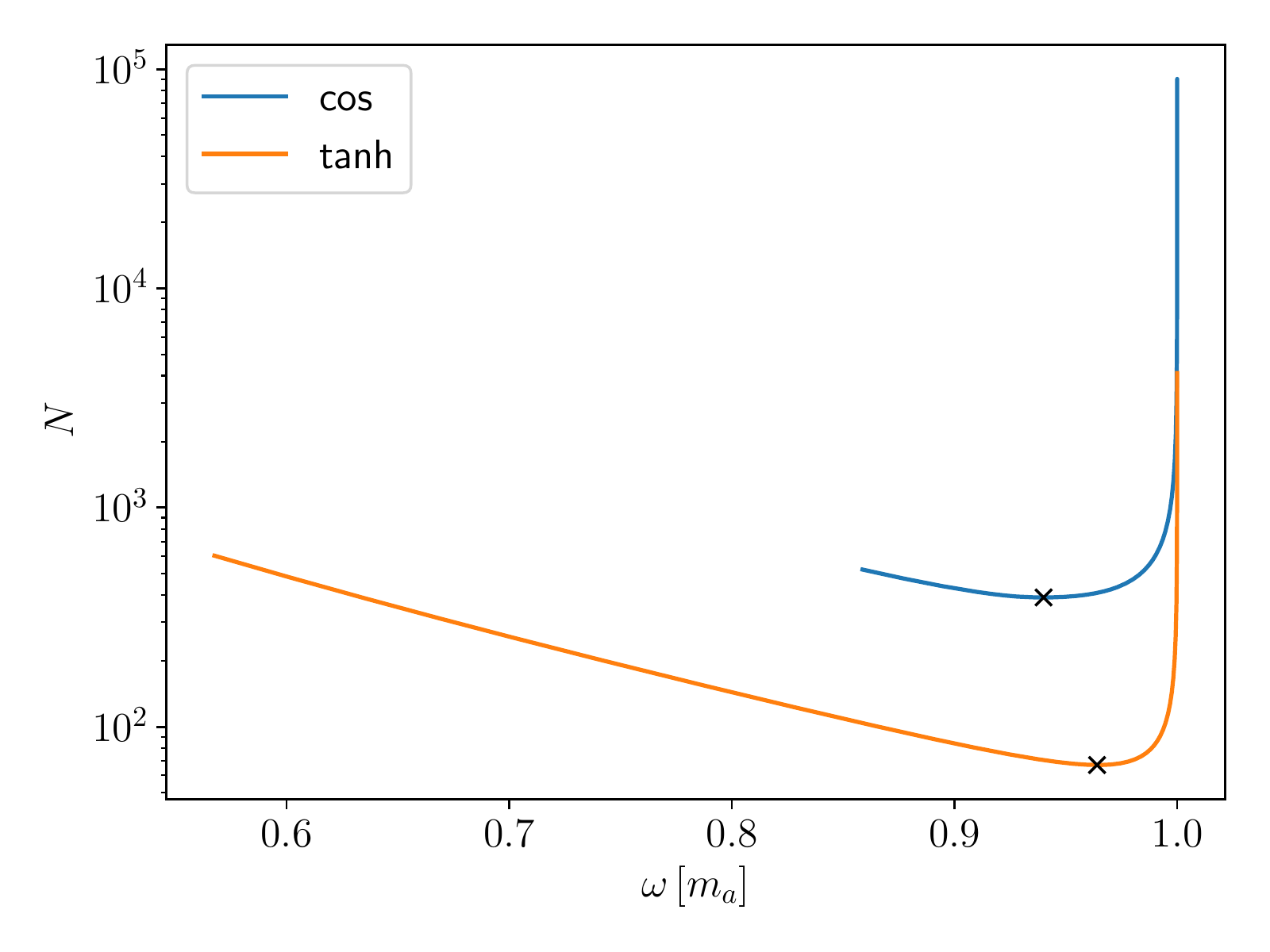}
	\caption{Particle number $N$ as defined by \eqref{N} as a function of the clump frequency $\omega$ for the cos- and tanh-potential. The leftmost point on each curve is where the central amplitude $\phia=5f_a$. For a given curve, the x marks the critical frequency for which the solutions becomes unstable \eqref{stability}}\label{part_num_of_omega}.
\end{figure}

For the axion clumps in question, the approximate axion particle number can be computed using \cite{Zhang:2020bec}
\begin{equation}
N \approx \frac{\omega}{2}\phia^2\int \dd^3 x\,f^2(r).\label{N}
\end{equation}
  For a clump to be stable towards long-wavelength perturbations, we need to satisfy the Vakhitov and Kolokolov stability condition \cite{Kolokolov1973,LEE1992251,Nugaev2020}
\begin{equation}
\frac{\dd N}{\dd\omega}<0.\label{stability}
\end{equation}
We denote the frequency for which this condition is satisfied as $\omega_{\text{crit}}$. Note that $\omega_{\text{crit}}<m_a$.
For all our analysis we initialize the clump solutions at some $\omega<\omega_{\text{crit}}$. The axions can in principle radiate through electromagnetic radiation or small scalar wave radiation which depletes the clump and increases the axion frequency untill it hits $\omega_{\text{crit}}$. Of course, in this paper we will mostly consider radiation via the EM sector. At the instant when the critical frequency is reached, the axion clump solution will collapse. Therefore, we limit our analysis to frequencies below $\omega_{\text{crit}}$. In figure \ref{part_num_of_omega} we plot the number of particles \eqref{N} as a function of frequency computed from the profiles obtained in the previous section. The lowest frequency shown for each potential corresponds to a central amplitude of $\phia = 5 f_a$. Following the criteria \eqref{stability} the clump becomes unstable and decays rapidly for $\omega>\omega_{\text{crit}}=0.94m_a \,(\phia<3.78 f_a)$ in the case of the cos-potential, while for the tanh-potential this corresponds to $\omega>\omega_{\text{crit}}=0.964m_a\,(\phia<1.06 f_a)$. This sets the lower bounds on the amplitudes $\phia$ we will consider for each potential. The frequency for which the clump becomes unstable is shown by a cross on each curve in figure \ref{part_num_of_omega}. Following a curve from lower to higher frequency (left to right), the solutions are stable until we pass the x, and are unstable afterwards. 
\section{electromagnetic radiation}\label{emag}
We now review the computation of electromagnetic radiation emanating from an axion clump oscillating in time. We assume the clump amplitude and frequency to be time independent. This amounts to ignoring back-reaction. We will allow the clump amplitude to have a spatial profile. More specifically, we will use the spatial profiles obtained in the previous section when we present the results for radiated electric and magnetic fields. 
We start by writing the magnetic and electric field in terms of a vector (scalar) potential $\A(\x,t)$ ($\Phi(\x,t)$) with 
\begin{align}
\Er(\x,t) &= -\pt \A(\x,t)-\nabla\Phi(\x,t),
\\
\Br(\x,t) &= \nabla\times \A(\x,t).
\end{align} 
Choosing the Lorenz gauge, the equations \eqref{dot(E)} and \eqref{dot(B)} then transform to 
\begin{align}
\square \A(\x,t) &= -\g\dot{\phi}(\x,t)\Bs\equiv \J(\x,t),\label{current}
\\
\square\Phi(\x,t) &= \g\nabla\phi(\x,t)\cdot\Bs\equiv \rho(\x,t).\label{charge}
\end{align}
Using the Green's function 
\begin{equation}
G(\x,t;\x',t') = -\frac{\delta(t-t'-|\x-\x'|)}{4\pi|\x-\x'|},\label{greens 1}
\end{equation}
we can then find the electromagnetic potentials $H(\x,t) = (\Phi(\x,t), \A(\x,t))$ given a source $S(\x,t) = (\J(\x,t),\rho(\x,t))$ by
\begin{equation}
H(\x,t) = \int\dd^4 x \,G(\x,t;\x',t')S(\x',t'). \label{greens eq}
\end{equation}
Given that we have an expression for the axion field $\phi(\x,t)$, we can then find the electromagnetic fields that solve \eqref{dot(E)} and \eqref{dot(B)}. The explicit expressions for the radiated electric and magnetic fields for the ansatz in Eq. \eqref{ansatz_2} are provided in \cite{Sen:2021mhf} and the corresponding expression for the radiated power is given by
\beq
P=\g^2\frac{\phia^2B_0^2\omega^3R^4\pi^5}{12 k_\omega}\left(\frac{\tanh(\pi k_{\omega} R/2)}{\cosh(\pi k_{\omega} R/2)}\right)^2
\label{power}
\eeq
where $k_\omega$ is the axion wavenumber with $k_{\omega}=\omega$ in vacuum (no plasma). The expression for radiated power changes when the clump is submerged in a plasma. In this case the matter current $J_m^{\mu}$ captures the effect of this plasma \cite{Sen:2018cjt},  where one can use linear response $\J_m=\sigma \E$ with $\sigma$, a frequency dependent conductivity. Moreover, one can use the Drude model to arrive at the frequency dependence of $\sigma$. As an example, consider the interstellar medium which consists of a plasma of ionized Hydrogen with free electrons scattering off of them. In that case, $\sigma$ depends on the collision time of the electrons in the plasma. In the collision-less limit, $\sigma$ is purely imaginary and the photon acquires a plasma mass.
The  Green's function changes \cite{Amin:2021tnq, Sen:2021mhf} to
\begin{align}
G(\x,t;\x',t')=&-\int \frac{d\omega}{2\pi}\frac{\theta\left(\omega^2-\omega_p^2\right)}{4\pi\norm{\x-\x'}}\Big(e^{i\sqrt{\omega^2-\omega_p^2}\norm{\x-\x'}-i\omega(t-t')}\theta(\omega)
\nonumber
\\
&\quad\quad\quad+e^{-i\sqrt{\omega^2-\omega_p^2}\norm{\x-\x'}-i\omega(t-t')}\theta(-\omega)\Big)\nonumber\\
&\quad\quad-\int \frac{d\omega}{2\pi}\frac{\theta\left(-\omega^2+\omega_p^2\right)}{4\pi\norm{\x-\x'}}e^{-\sqrt{|\omega^2-\omega_p^2|}|\x-\x'|-i\omega(t-t')},\label{greens 2}
\end{align}
where $\op$ is the plasma frequency (mass) and $\theta(x)$ is the Heaviside step function. The expression for radiated power can be obtained by replacing $k_{\omega}$ by $\sqrt{\omega^2-\omega_P^2}$ in the expression of Eq. \ref{power} and also by multiplying with a unit step function $\theta(k_{\omega})$ to ensure that there is no radiation below the plasma frequency. With these expressions we now have all the ingredients to implement the back-reaction analysis. Before we do so, let us discuss 
a few interesting applications where back-reaction can reveal new insights about the evolution of the axion clump. 
\section{Applications}\label{applications}
When considering the radiated power from axion clump as shown in Eq. \ref{power} there are several interesting features to note. For example, the radiation
is a function of the size of the clump $R$ in such a way that it peaks approximately when the hyperbolic function in the brackets of \eqref{power} is maximized, i.e. at $\pi k_{\omega} R/2=\log(\sqrt{2}+1)$, and is suppressed for both $\pi k_{\omega}R/2\ll 1$ and $\pi k_{\omega}R/2\gg 1$.  However, in a system where $\omega$ itself is evolving with time, as is the case in a system with back-reaction, one can have a clump that radiates efficiently at some instant $t=t_0$ with $\pi k_{\omega} R/2\sim \log(\sqrt{2}+1)$, while at a later time $t=t_1>t_0$ the radiation gets suppressed as $\omega$ increases and $\pi k_{\omega} R/2\gg \log(\sqrt{2}+1)$. 

\begin{figure}
	\begin{subfigure}{0.49\textwidth}
		\includegraphics[width=\textwidth]{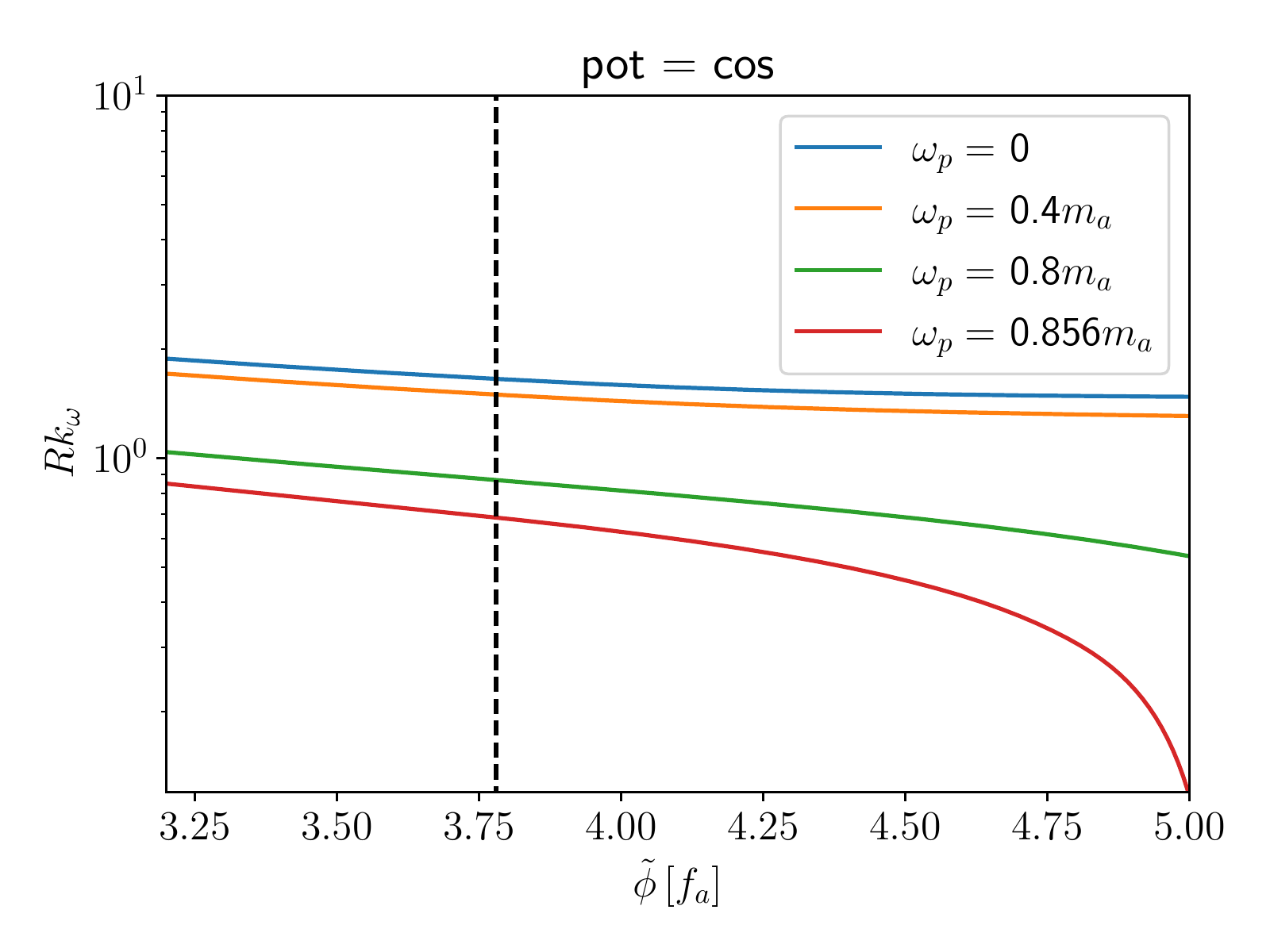}
	\end{subfigure}
	\hfill
	\begin{subfigure}{0.49\textwidth}
		\includegraphics[width=\textwidth]{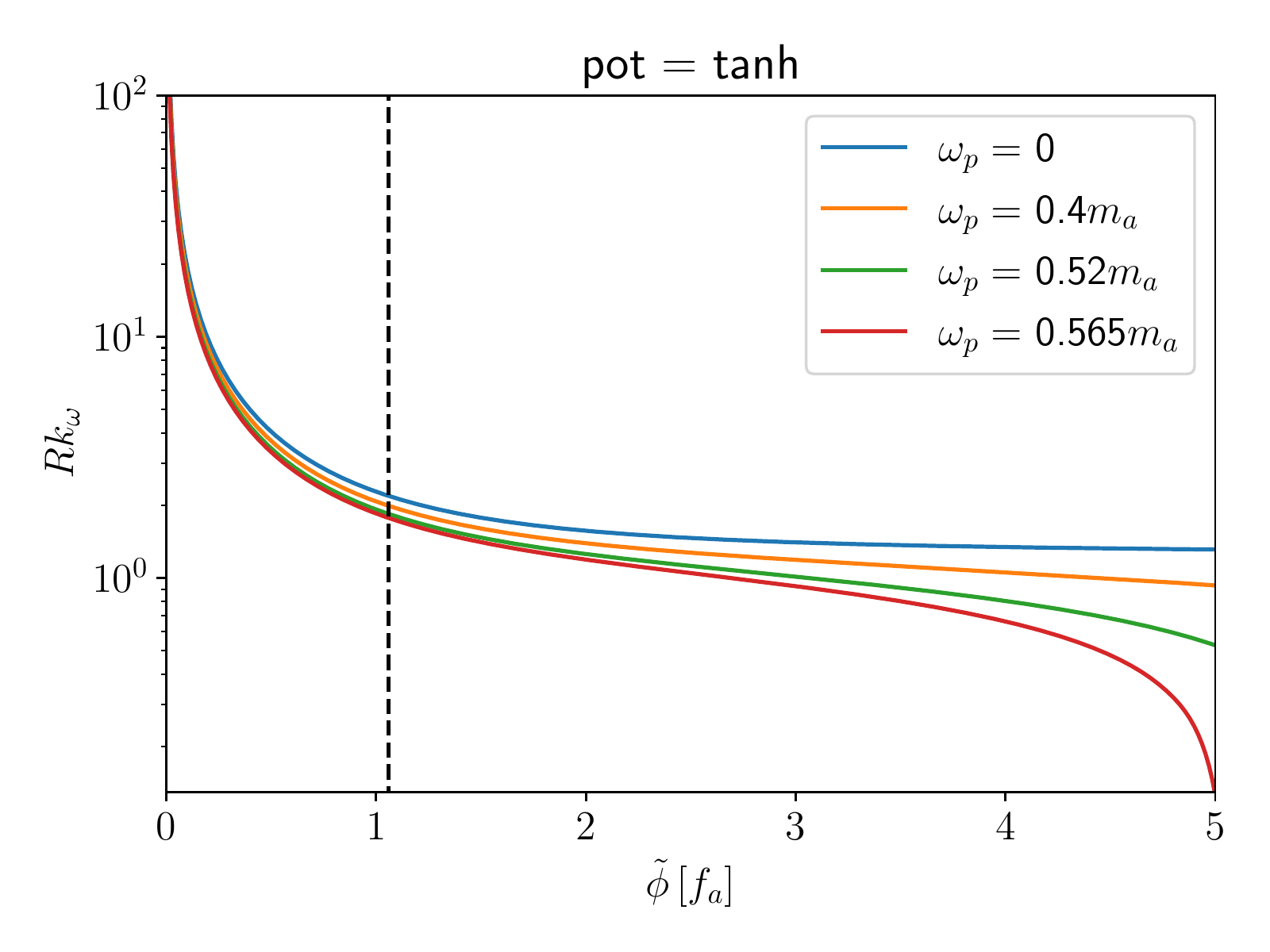}
	\end{subfigure}
	\caption{The product $R k_\omega $ for a few values of the plasma frequency in the case of the two potentials considered. The vertical line represents the central amplitude at which $\omega=\omega_\text{crit}$.}\label{R_k_of_phi0}
\end{figure}

In a similar manner, a clump that was practically non-radiating via the EM sector at a certain instant in time can start to radiate EM waves efficiently at a later instant. For example, consider an axion clump in the presence of a static external magnetic field and submerged in a plasma with the plasma mass being close to the axion mass. In a plasma we have $k_\omega = \sqrt{\omega^2-\op^2}$. Let's assume that, $\omega>\omega_P$ and $\pi k_{\omega}(t)R(t)/2\ll 1$ at an instant $t=t_0$. Therefore, at time $t=t_0$ there is a small amount of electromagnetic radiation, which depletes the clump slowly and raises its frequency in the process. Consequently, at a later time $t=t_1$, we may find $\pi k_{\omega}(t)R(t)/2\sim \log(\sqrt{2}+1)$ due to the increase in frequency which allows the clump to radiate EM waves efficiently. 
In figure \ref{R_k_of_phi0} we plot the product $k_\omega R$ as a function of the central amplitude $\phia$ to illustrate this point. 
We have highlighted the central amplitude where the long wavelength instability kicks in $\phia = \phia(\omega_{\text{crit}})$ by the vertical dashed lines. 
The change in the radiation intensity described above cannot be captured unless one includes back-reaction in the analysis. 

Note that, we can also imagine a slightly modified scenario. Consider again an axion clump in a plasma and a static magnetic field. We take $\omega_P\neq 0$, $k_{\omega_{\text{crit}}}^2>0$, where at an initial instant in time $t=t_0$, $k_{\omega}^2<0$. In this case power radiated through EM waves at earlier times is zero. Therefore, it would appear that the clump frequency doesn't change with time and the clump never achieves the resonant radiation condition. However, it is important to remember that the scalar radiation emanating from the clump will deplete the clump even when EM radiation is zero. We ignore the scalar radiation in the other examples in this paper because our parameters are chosen such that EM radiation typically dominates over scalar radiation. However, in this particular example, at early times, this is not the case and scalar radiation has to be taken into account to describe the dynamics of the clump. Because of the scalar radiation, the clump will change its frequency and will achieve $k_{\omega}^2>0$ at a later time causing a sudden burst of EM radiation and eventually reaching the resonant condition. In section \ref{results} we will demonstrate this behavior of axion clumps by combining results from \cite{Zhang:2020bec} with ours. Note that, although we don't analyze scalar radiation explicitly in this paper, it is simple to combine the scalar radiation results from \cite{Zhang:2020bec} with our results since in earlier times where scalar radiation is crucial, EM radiation is completely absent. This allows us to borrow the results of \cite{Zhang:2020bec} for early times. Similarly, later, when EM radiation takes over, scalar radiation is negligible and the framework for EM radiation developed in this paper holds.

Another interesting situation can arise when an axion clump submerged in a plasma is subjected to an external magnetic field which itself has an alternating time dependence with frequency $\Omega$. Of course, for $\Omega$ to be relevant to the discussion of radiation and axion evolution, it has to be of the order of the axion mass. This can be the case for ultralight axions in astrophysical magnetic fields: for example magnetic fields of rapidly rotating pulsar can have frequencies of the order of $10^{-11}$eV making them relevant for ultralight axions of the same mass scale. 
In a surprising coincidence, the plasma frequency of the interstellar medium is also in the range of $10^{-11}-10^{-12}$ eV making it interesting for the kind of ultralight axions which can be sensitive to pulsar magnetic fields. The radiation for this case was analyzed in \cite{Sen:2021mhf}
where it was shown that the radiated power has contributions from two different frequencies $\omega+\Omega$ and $\omega-\Omega$. The corresponding expression for the radiated power is given by
\begin{align}
\langle P\rangle\approx&\frac{4\pi}{3}\left(\frac{C\beta}{\pi f_a}\frac{B_0 \omega \phia}{8}\pi^2R^2\right)^2
\nonumber
\\
&\times\left[\frac{(\omega+\Omega)}{ k_{\omega+\Omega}}
\frac{\tanh\left(\frac{\pi k_{\omega+\Omega}R}{2}\right)^2}{\cosh\left(\frac{\pi k_{\omega+\Omega}R}{2}\right)^2}\theta( k_{\omega+\Omega}^2)+\frac{\norm{\omega-\Omega}}{ k_{\omega-\Omega}}\frac{\tanh\left(\frac{\pi k_{\omega-\Omega}R}{2}\right)^2}{\cosh\left(\frac{\pi k_{\omega-\Omega}R}{2}\right)^2}\theta( k_{\omega-\Omega}^2)
\right],\label{P}
\end{align}
 where $k_{\omega}=\sqrt{\omega^2-\op^2}$. 
 
Consider a scenario where at some initial time $t_0$, $\omega(t_0)>\Omega$ and $|\omega_{\text{crit}}-\Omega|>\omega_P$. However, the clump is initialized such that $|\omega(t_0)-\Omega|<\omega_P$ at a certain instant in time. In that case the second term in the expression in Eq. \ref{P} doesn't contribute to any radiation since the Heaviside function sets it to zero. The first term on the RHS of course contributes to radiation. As the clump keeps radiating due to the first term, $\omega$ keeps increasing with time approaching $m_a$. At some point in time $\omega$ becomes large enough that $|\omega-\Omega|=\omega_P$ which turns on the second term. At this instant the radiated power will exhibit a rapid increase. We will demonstrate this explicitly in section \ref{results}. There is something even more interesting that can happen while considering the radiated power in Eq. \ref{P}. Imagine almost the same scenario with $\omega_{\text{crit}}>\omega>\Omega>0$ and $|\omega-\Omega|>\op$ at some initial time $t_0$. Let's also assume that 
$\pi |k_{\omega_{\text{crit}}-\Omega}|R/2\gg 1$. The clump is initialized such that $\pi k_{\omega-\Omega}R/2\ll 1$ and $\pi k_{\omega+\Omega}R/2\gg1$ with $\omega_{\text{crit}}>\omega>\Omega$. In this case both the first and the second terms in Eq. \ref{P} radiate at a moderate rate since both are outside of resonance. This causes $\omega$ to increase such that at a certain instant the condition $\pi k_{\omega-\Omega} R/2\sim \log(\sqrt{2}+1)$ is satisfied placing the second term in resonance. This leads to a rapid increase in radiation. Section \ref{results} will also contain explicit demonstration of this behavior. 



\subsection{Strategy for computing back-reaction}\label{power radiated}
We will now put together the two pieces of the calculation from section \ref{axion profile section} and \ref{emag} in order to account for back-reaction. These are the axion profile solutions and electromagnetic radiation obtained for those profiles. Our strategy will be to initiate the clump at time $t=t_0$ with some central amplitude $\phia$, which is taken to be an average of the central amplitude over a time period of the radiation as outlined in section \ref{argument}. This fixes the value for $\omega$ and $R$ (both averaged over one time period) for the corresponding clump solution so as to satisfy Eq. \ref{shooting}. It also fixes the average particle number and the average total energy for the axion clump configuration. We will omit the usage of the word average from this point onwards. We compute the total energy $E(t=t_0)$ of this configuration, as well as the radiated power $P(t=t_0)$, using the results quoted in the previous section. A short time interval $\Delta t>2\pi/\omega(t=t_0)$ later, the clump energy will have changed to $E(t=t_0+\Delta t) = E(t=t_0)-\Delta t P(t=t_0)$. Note that the condition $\Delta t>2\pi/\omega$ is important in order for this strategy to make sense. Similarly, $\Delta t$ has to be chosen in such a way that it's much smaller than the EM decay timescale. As mentioned earlier, in the limit of $\gp\ll 1$, the decay time of the clump $\tau_{\text{EM}}$ is large compared to the time period of radiation which allows for a range of values for $\Delta t$ where $\tau_{\text{EM}}>\Delta t>\frac{2\pi}{\omega}$ . Just as argued in section \ref{argument}, as energy is radiated away via propagating electromagnetic fields, the clump reconfigures itself into a solution of the sourceless profile equation \eqref{sourceless_eq} over the time step $\Delta t$. The updated energy will then give a new (smaller) value for the central amplitude $\phia$ and thus a new set of values for $\omega$ and $R$ at time $\Delta t$. We continue this cycle until we reach the critical frequency $\omega_{\text{crit}}$, defined as  
\begin{equation}
\frac{\dd N}{\dd \omega}\Big|_{\omega = \omega_{\text{crit}}} = 0,\label{crit}
\end{equation}
which is the frequency where the long wavelength instability kicks in and the clump collapses.

\begin{figure}
	\begin{subfigure}{0.49\textwidth}
		\includegraphics[width=\textwidth]{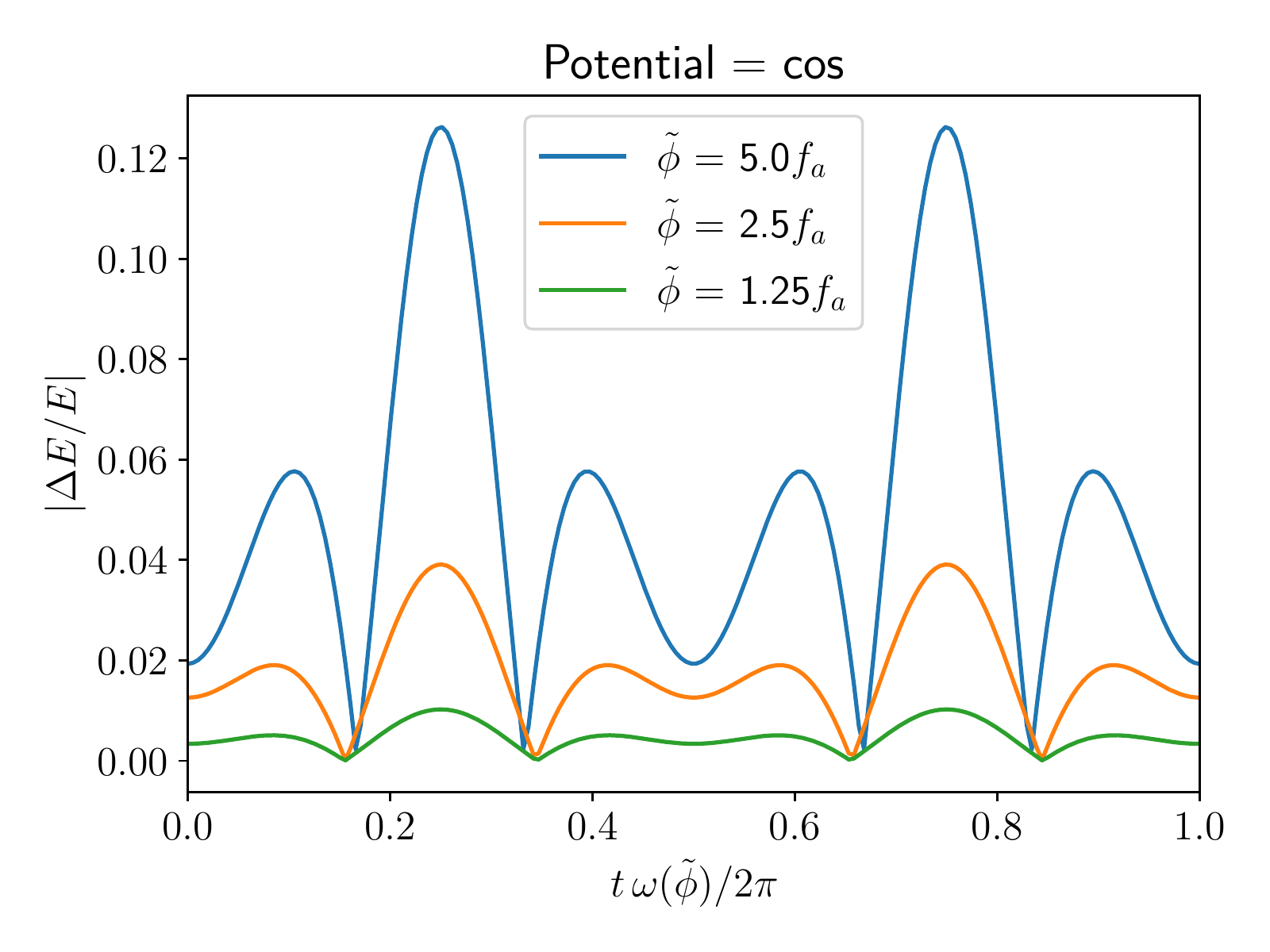}
	\end{subfigure}
	\hfill
	\begin{subfigure}{0.49\textwidth}
		\includegraphics[width=\textwidth]{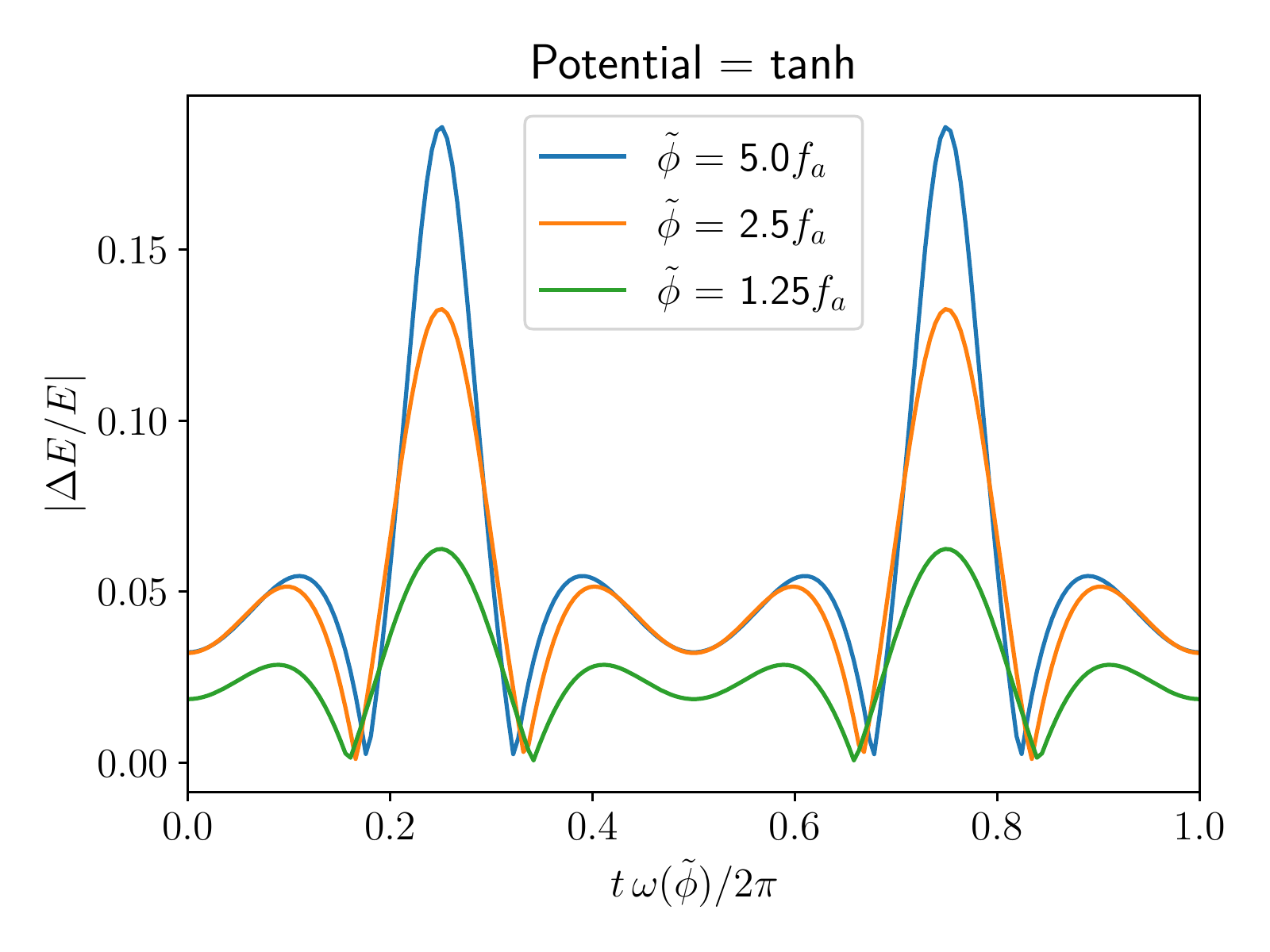}
	\end{subfigure}
	\caption{The figure shows the variation in energy $E$ computed using Eq. \ref{E} and its time average $\langle E\rangle$ computed using \ref{av} for the ansatz in Eq. \ref{ansatz}. $|\frac{\Delta E}{E}|$ with $\Delta E = E-\langle E\rangle$ is plotted for a few different values of the central amplitude $\phia$.}\label{Energy_variance}
\end{figure}

As mentioned in the previous paragraph, when dealing with the clump amplitude $\phia$, its frequency $\omega$ and size $R$ etc, we are really referring to their values averaged over one time period of the radiation. This averaging should in principle be inconsequential for the total energy. This is because for an exact stable solution of the EOM, the total enegy should be constant in time. However, while computing the solutions to Eq. \ref{sourceless_eq} we had used a single frequency ansatz dominating the clump time variation. This is an approximate solution. As a result, the energy is approximately constant over one time period and needs to be averaged just like the other quantities. The expression for the averaged energy can be obtained from 
\begin{equation}
E = \int\dd^3 x\, \Big[\frac{1}{2}\dot{\phi}^2+\frac{1}{2}(\nabla\phi)^2+V(\phi)\Big]
\label{E}
\end{equation}
to be
\begin{equation}
\langle E\rangle = \int\dd^3 x\,\Big[\frac{1}{4}\phia^2\omega^2\phir(r)^2+\frac{1}{4}\phia^2(\nabla\phir(r))^2+\langle V(\phi)\rangle\Big],
\label{av}
\end{equation}
where 
\begin{equation}
\langle V(\phi)\rangle \equiv \frac{\omega}{2\pi}\int_{-\pi/\omega}^{\pi/\omega}\dd t'\, V(\phia\phir(r)\cos(\omega t'))
\end{equation}
and $\phia$, $\omega$ and $R$ are the time-averaged central amplitude, frequency and size. 
In figure \ref{Energy_variance} we plot the fluctuations in the energy as a function of time for a few different values of the central amplitude $\phia$, over a period of the clump. As expected, the single frequency approximation becomes better for small values of $\phia$, since the higher order contributions from the axion potential gets suppressed as $\phia$ decreases.

\subsection{Results}\label{results}
\subsubsection{Radiation with time}\label{radiation_with_time}
We will now see how the introduction of back-reaction effects the clump evolution as well as the evolution of the radiated power. This will also reveal the timescale over which a clump depletes as it radiates.
Following the procedure outlined in the previous section, we initiate the axion clump at $\phia = 5f_a$. This gives us an initial frequency of $\omega = 0.856m_a (0.568m_a)$ as well as a clump size $R = 1.75m_a^{-1} (2.34m_a^{-1})$ for the cos- (tanh-) potential. Once these parameters are determined, we immediately also have the initial energy $E(t=t_0)$. Anticipating $\tau_{\text{EM}}\sim \frac{1}{m_a}\left(\frac{\pi m_a f_a}{C\beta B_0}\right)^2$, we choose $\Delta t=\frac{10^{-3}}{m_a}\left(\frac{\pi m_a f_a}{C\beta B_0}\right)^2>2\pi/\omega$. The results are not significantly altered if we change $\Delta t$ by an order of magnitude. With this we can now apply the algorithm outlined in the previous section. 

\begin{figure}
	\includegraphics[width=\textwidth]{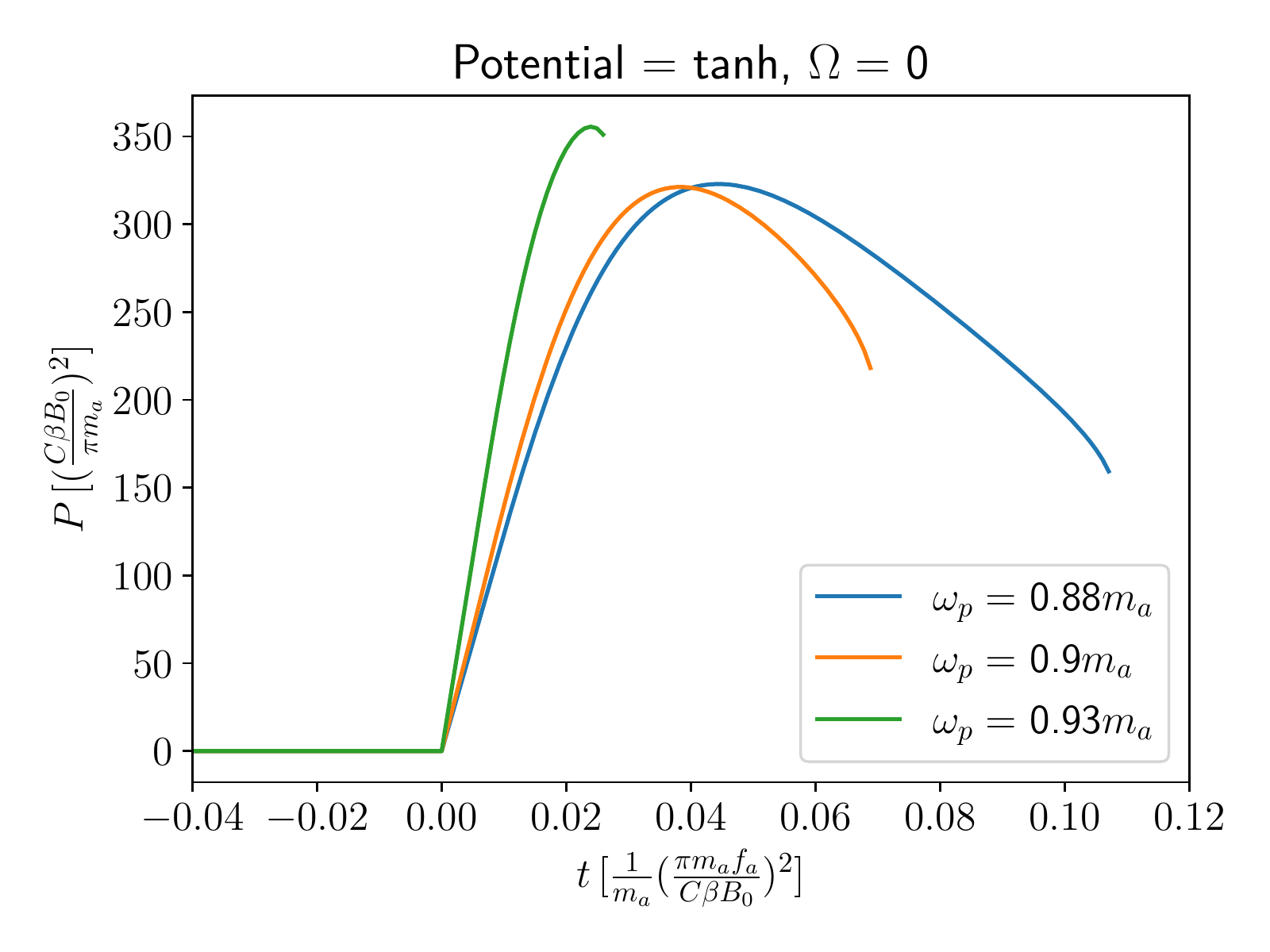}
	\caption{The electromagnetic radiated power as a function of time in the case of an axion clump that is initiated with a clump frequency $\omega(t_0)$, less than the plasma frequency. Time $t=0$ is defined as the time when $k_\omega = 0$, while for $t<0$ the clump only evolves through the emission of scalar waves. The rightmost point on each curve is the time at which $\omega = \omega_\text{crit}$}.\label{P_of_time_scalar}
\end{figure}

We begin with a clump submerged in a plasma. We take the external magnetic field to be completely static such that $\Omega=0$. We also take the axion potential to be the tanh potential. In Fig \ref{P_of_time_scalar} we initialize the clump at $t=t_0<0$ with a frequency $\omega<\omega_P<\omega_\text{crit}$. In this case, there is no EM radiation at early times as seen from Fig. \ref{P_of_time_scalar}. During this time, the clump depletes through scalar radiation changing its frequency. At some later time, which we have set to be $t=0$ for convenience, the clump achieves $k_\omega^2>0$. At this point the clumps starts to radiate EM waves and we see a corresponding rise in EM power in Fig. \ref{P_of_time_scalar}. The radiated power increases quickly as $\omega$ increases towards the resonance $\pi R  k_\omega/2\approx \text{log}(\sqrt{2}+1)\approx 0.88$. However, the rapid decrease in the central amplitude $\phia$ somewhat curtails the radiated power before the resonance is reached. Yet, we see how a dynamical change in clump frequency via scalar radiation can turn a clump that is not emitting any EM radiation to a clump that does so. 
In the plot, we have set the rightmost point on each curve to represent the time at which the clump frequency reaches $\omega=\omega_{\text{crit}}$. Note that we see in figure \ref{P_of_time_scalar} that clumps that are initiated inside a plasma with a lower plasma frequency emit electromagnetic radiation over a longer time period. This is  explained in the following way: for a high plasma frequency, most of the clump must be depleted in the form of scalar waves before we can reach the condition $k_\omega=0$ at which point the clump begins to radiate via EM waves. For a plasma frequency that is lower, which as a result is closer to the initial clump frequency, the clump has to radiate via scalar waves for less time to reach the condition $k_\omega=0$. In other words, for lower plasma frequency, there is simply more left of the clump to radiate away at the time the condition $k_\omega=0$ is met. 

We now consider an axion clump submerged in a plasma of plasma frequency $\omega_P$ subjected to an external alternating magnetic field of frequency $\Omega$. 
In figure \ref{P_of_time_cos} and \ref{P_of_time_tanh} we show the corresponding radiated power while taking into account back-reaction for the two potentials (cos and tanh)  for a few different values of $\Omega$ and $\op$. The clump in each case is initiated with $\phia(t_0) = 5f_a$, and we have adjusted the time axis in such a way that $\omega(t=0) = \omega_\text{crit}$. The left most point on the time-axis for each curve then corresponds to the initial time $t_0<0$ with $\phia(t_0)=5f_a$ (for example, the red curve in the left panel in figure \ref{P_of_time_cos} is initiated with $t_0 \approx -0.38\frac{1}{m_a}\left(\frac{\pi m_a f_a}{C\beta B_0}\right)^2$, and as time evolves we eventually reach $\omega(t=0) = \omega_\text{crit}$). 

Let us first consider the case of the cosine potential which is shown in Fig. \ref{P_of_time_cos} in two panels, the left corresponds to $\Omega=0.1 m_a$ and the right $\Omega=0.55 m_a$. In this case we set $\omega(t_0) = 0.8586m_a$ and $R(t_0)=1.476m_a^{-1}$. Note that, $t_0$ for the different plasma frequencies will be different as seen in the plots. For $\op=0$ with $\Omega = 0.1m_a$ as shown in the left panel of figure \ref{P_of_time_cos} we have that $\pi k_{\omega+\Omega} R/2$ = 2.22 and $\pi R k_{\omega-\Omega}/2$ = 1.759 when $t=t_0$. This puts both terms in equation \eqref{P} just outside the resonance which occurs at $\pi k R/2 =  0.88 $, and since $k_{\omega\pm\Omega}$ increases with time as $\omega$ increases, we fall further and further away from the resonance as time passes. In addition since $\phia$ decreases with time, the factor in front of the brackets in \eqref{P} decreases with time as well. As a result, the radiation decreases monotonically with time.

The case $\Omega = 0.55 m_a$, with $\op=0$ is shown in the right panel of figure \ref{P_of_time_cos}. For these parameters, the first term in equation \eqref{P} is outside of the resonance with $\pi  R(t_0) k_{\omega+\Omega}(t_0)/2 >0.88$ whereas the second term is also out of resonance, albeit with, with $\pi  R(t_0) k_{\omega-\Omega}(t_0) = 0.71<0.88$. As time progresses, the frequency increases and the first term moves further away from the resonance while the second term gets closer to it. This change should contribute to an increase in radiated power. But we still get an overall decrease in radiated energy with time as the amplitude of the clump decreases faster compensating for any increase in radiation due to the resonance.

The case of a plasma frequency of $\op = 0.5m_a$ with $\Omega = 0.1m_a$ and $\op = 0.2m_a$ with $\Omega = 0.55m_a$ are quite similar to that of $\omega_P=0$, with the main difference being that the factor $\frac{|\omega\pm\Omega|}{k_{\omega\pm\Omega}}>1$ when $\op\neq 0$, explaining an overall increase in radiated power compared to $\op=0$.

A quite different scenario emerges once the plasma frequency is large enough so that $|\omega_{\text{crit}}-\Omega|>\op>|\omega(t_0)-\Omega|$. In this case, at the initial time $t=t_0$ the second term in equation \eqref{P} is zero as $k_{\omega-\Omega}^2(t=t_0)<0$. The first term in Eq. \eqref{P} radiates at a moderate rate. This causes $\omega$ to increase with time and at a later time $t_1$, $k_{\omega-\Omega}^2(t=t_1)>0$, giving a sudden increase in the radiated power. In contrast with the case where $\Omega=0$ and $\omega_P\neq 0$, in this case we don't need assistance from scalar radiation to alter the frequency of the clump. We see this effect for the green and red curves in figure \ref{P_of_time_cos}. This happens for both $\Omega = 0.1m_a$ and $\Omega = 0.55m_a$, however, for $\Omega = 0.55 m_a$ the overall radiation is smaller at earlier times, since the first term in equation \eqref{P} is further out of the resonance than in the case $\Omega = 0.1m_a$. 

Having looked at the case of a cos potential, we now turn to look at the radiated power in the case of the tanh potential, as shown in figure \ref{P_of_time_tanh}. The effects we see are similar to that of the cos case. At $t=t_0$ we have set $\omega(t=t_0) = 0.5676m_a$ and $R(t=t_0) = 1.954m_a^{-1}$ for $\phia(t_0)=5f_a$. For both blue and orange curves ($\op=0, 0.3 m_a$) we have $k_{\omega\pm\Omega}^2>0$ at all times and  $\pi k_{\omega\pm\Omega} R/2> \log(\sqrt{2}+1)$ at $t=t_0$. Therefore, both terms in equation \eqref{P} contribute to the radiated power and both lie just outside of resonance. As the frequency increases, both move further away from the resonance which leads to a monotonic decrease in the radiated power. On the other hand, when $\Omega = 0.5m_a$, for the blue and orange curves with $\op=0, 0.05 m_a$, we have a distinct increase in radiation at earlier times that later gets curtailed as the clump depletes. For example, consider $\op=0$ with $\Omega = 0.5m_a$, where we have $\pi k_{\omega-\Omega}(t=t_0) R(t=t_0)/2 = 0.2<\log(\sqrt{2}+1)\approx0.88$, which is slightly outside the resonance. Similarly, $\pi k_{\omega+\Omega}(t=t_0) R(t=t_0)/2 = 3.27>\log(\sqrt{2}+1)\approx 0.88$, which makes the first term in Eq. \ref{P} outside of the resonance as well. With increasing $\omega$ the second term in Eq. \ref{P} moves closer to resonance leading to enhancement in radiation, until at some point it moves out of the resonance. The non-monotonic behavior of radiation for $\Omega=0.5 m_a$ with $\op=0.05 m_a$ (orange curve) can be explained in a similar way. 

Finally, we have the case where $k_{\omega-\Omega}^2(t=t_0)<0$, as shown by the red and green lines in figure \ref{P_of_time_tanh}. In this case, at $t=t_0$, the second term in Eq. \eqref{P} doesn't contribute to any radiation. For $\Omega = 0.1m_a$, we have the first term in equation \eqref{P} radiating significantly at earlier times, then falling out of resonance as $\omega$ increases with time. At some later time $t_1$ we have that $k_{\omega-\Omega}^2(t_1)>0$, which gives rise to the radiation peaks we see on the left panel of figure \ref{P_of_time_tanh}. The same happens for the red and green curves for $\Omega=0.5 m_a$. However, this time the first term in equation \eqref{P} is far out of resonance, which is why the overall radiation is small at earlier times as compared to $\Omega=0.1 m_a$.  

It is clear from Fig. \ref{P_of_time_cos} and \ref{P_of_time_tanh} that the back-reaction of the axion clump plays a key role in how the EM radiation intensity changes over time. These effects are significantly more pronounced when a clump moves in or out of resonance as time passes. In the next few paragraphs we will analyze the decay rate of the axion clump taking into account back-reaction. As we expect, we will see that enhancement in radiation due to resonance will significantly impact the decay time of the clump.
\subsubsection{Clump decay with time} 
To understand the decay of clumps with time let's first consider the axion clump evolution corresponding to the scenario described in Fig. \ref{P_of_time_scalar}. In other words, the axion clump is submerged in a plasma $\omega_P\neq 0$ in the presence of a static external magnetic field $\Omega=0$. The plasma frequency is set to be $\op>\omega(t_0)$ such that the electromagnetic radiation is strictly zero at early times including $t=t_0$. As stated before, in early times for this case we need to take into account scalar radiation to see how the clump properties change with time. 
The clump at early times depletes via scalar radiation. We depict this in Fig. \ref{Compare_scalar_EM}. Then, as the clump decays, eventually the frequency rises to a point where $\omega(t=0)=\op$. At this point, the electromagnetic radiation turns on, and increases rapidly as we saw in figure \ref{P_of_time_scalar}. Once EM radiation starts, the power radiated via EM radiation dominates over the power radiated through scalar waves. This of course is consistent with the perturbative approach of our calculation which relies on EM decay time scale being smaller than scalar decay time scale. 
The EM radiation therefore causes the clump to deplete much faster than the scalar radiation, as seen from Fig. \ref{Compare_scalar_EM}. In Fig. \ref{Compare_scalar_EM} we show clump decay for three different values of the plasma frequency. In the figure we define $t=0$ to be where $\omega = \op$ for each case. The dashed line shows the time dependence of the clump amplitude if it decayed only through scalar radiation. Similarly, the solid line for $t>0$ shows how the clump evolves while emitting EM radiation alone. To a good approximation the evolution of $\phia(t)$ follows the black dashed curve for $t<0$ and the solid blue curve for $t>0$. For the scalar radiation, we have loosely read the data points of figure 4 in \cite{Zhang:2020bec}. In \cite{Zhang:2020bec} it was found that the decay timescale due to scalar radiation for the tanh-potential is $\tau_{\text{scalar}}\sim 10^6m_a^{-1}$. We anticipate the EM decay timescale due to electromagnetic radiation to be similar to the estimate \eqref{tau}, $\tau_\text{EM}\sim \frac{1}{m_a}\Big(\frac{\pi m_a f_a}{C\beta B_0}\Big)^2$. Thus, in order for our approach to be valid we need $\Big(\frac{\pi m_a f_a}{C\beta B_0}\Big)^2<10^6$. On the other hand, as discussed in section \ref{radiation_with_time}, we are also required to make sure that the time step $\Delta t$ satisfy the condition $\Delta t>2\pi/\omega(t)$ where $\omega(t)$ is of the order of $m_a$. Our choice of $\Delta t$ from sec. \ref{radiation_with_time} is $\frac{10^{-3}}{m_a}\Big(\frac{\pi m_a f_a}{C\beta B_0}\Big)^2$. In order to satisfy these constraints we need $10^6>\Big(\frac{\pi m_a f_a}{C\beta B_0}\Big)^2>10$. We use $\Big(\frac{\pi m_a f_a}{C\beta B_0}\Big)^2=3\cdot 10^4$ in Fig. \ref{Compare_scalar_EM} which safely satisfies the constraints. 

\begin{figure}
	\begin{subfigure}{0.49\textwidth}
		\includegraphics[width=\textwidth]{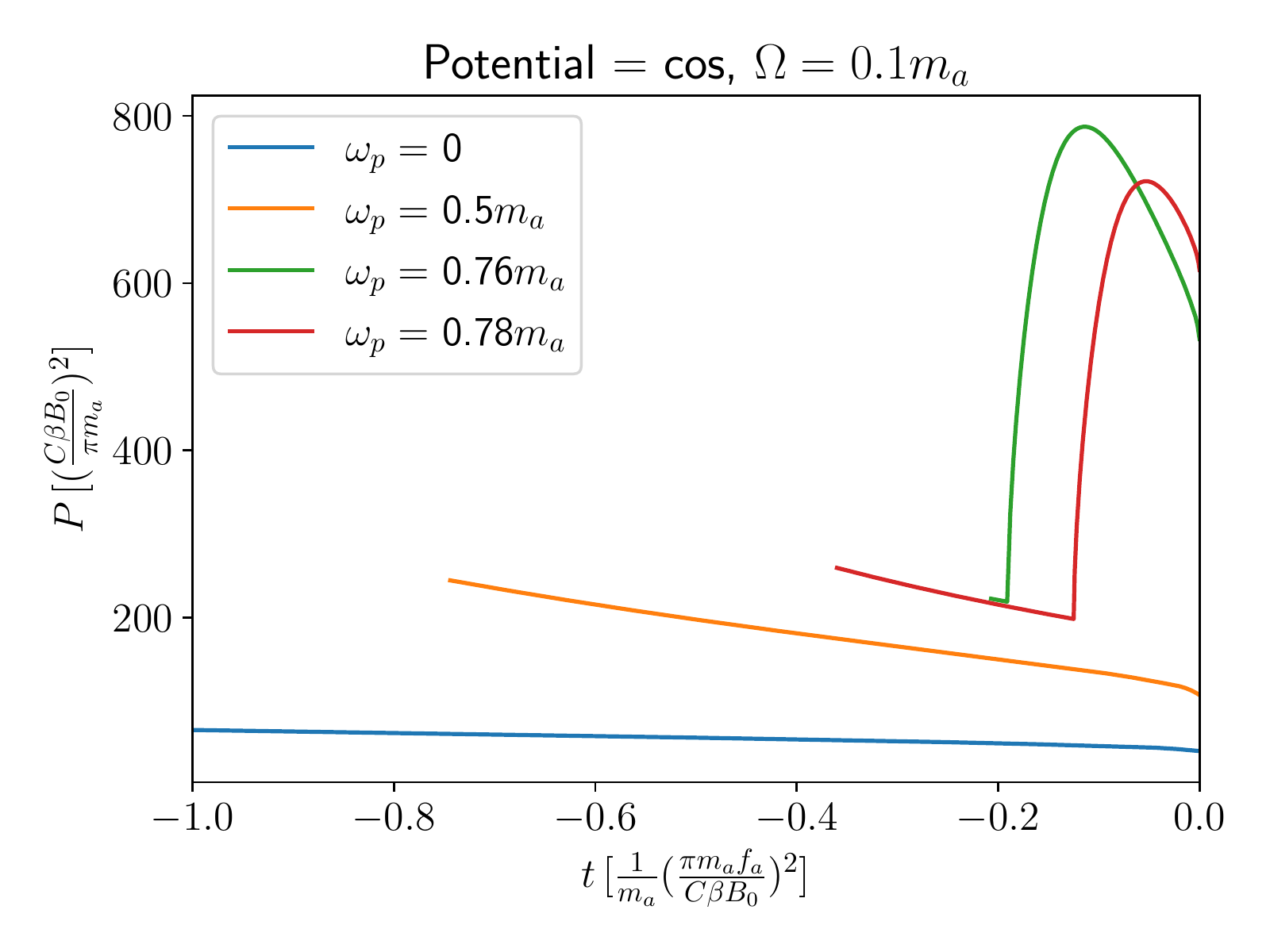}
	\end{subfigure}
	\hfill
	\begin{subfigure}{0.49\textwidth}
		\includegraphics[width=\textwidth]{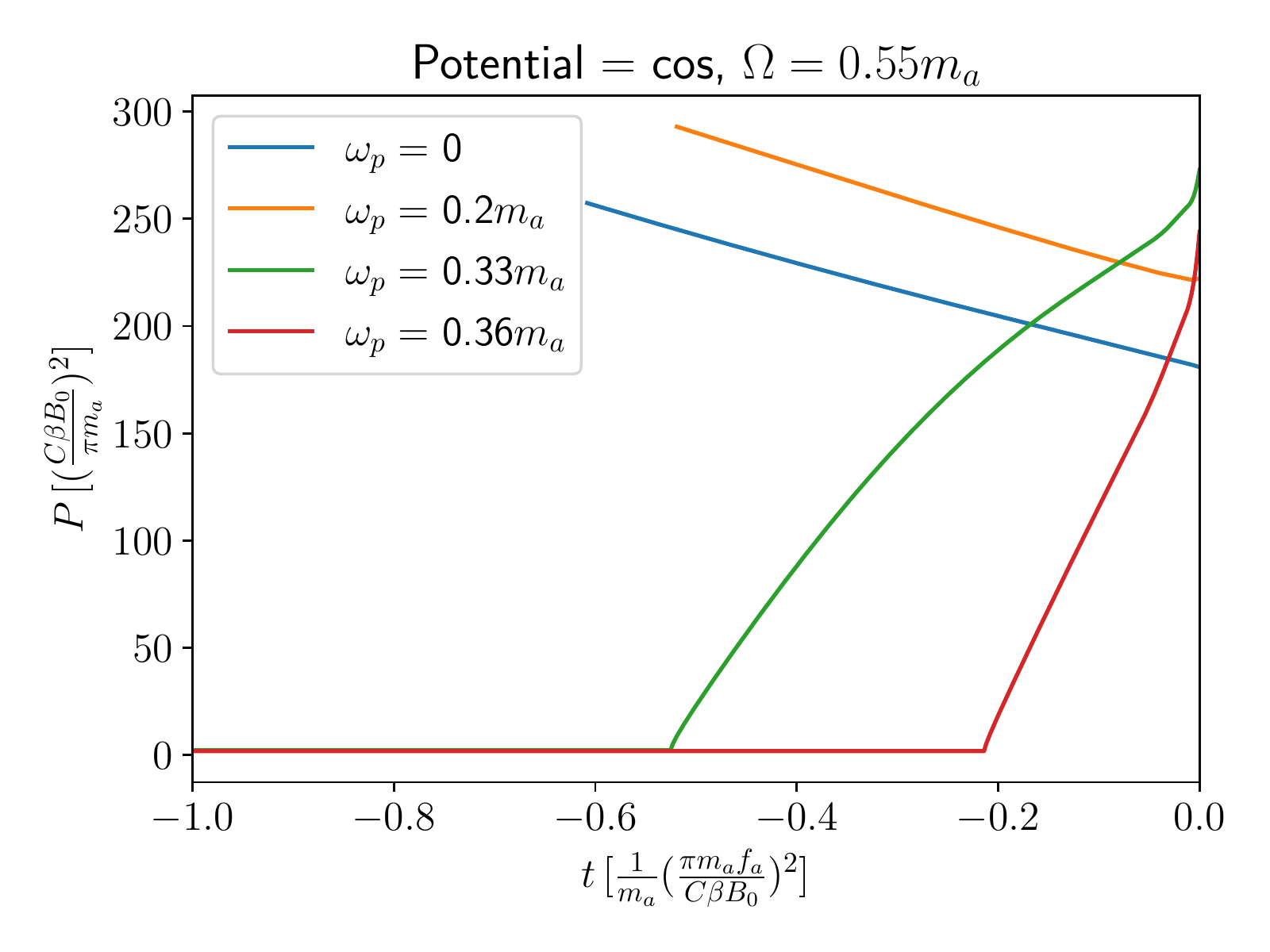}
	\end{subfigure}
	\caption{Power radiated as a function of time for the cos-potential for a few different values of the external magnetic field frequency $\Omega$ and the plasma frequency $\op$. We have defined $t=0$ to be the time at which $\omega(t=0) = \omega_\text{crit}$, where $\omega_\text{crit}$ is given by} \eqref{crit}.\label{P_of_time_cos}
\end{figure}

\begin{figure}
	\begin{subfigure}{0.49\textwidth}
		\includegraphics[width=\textwidth]{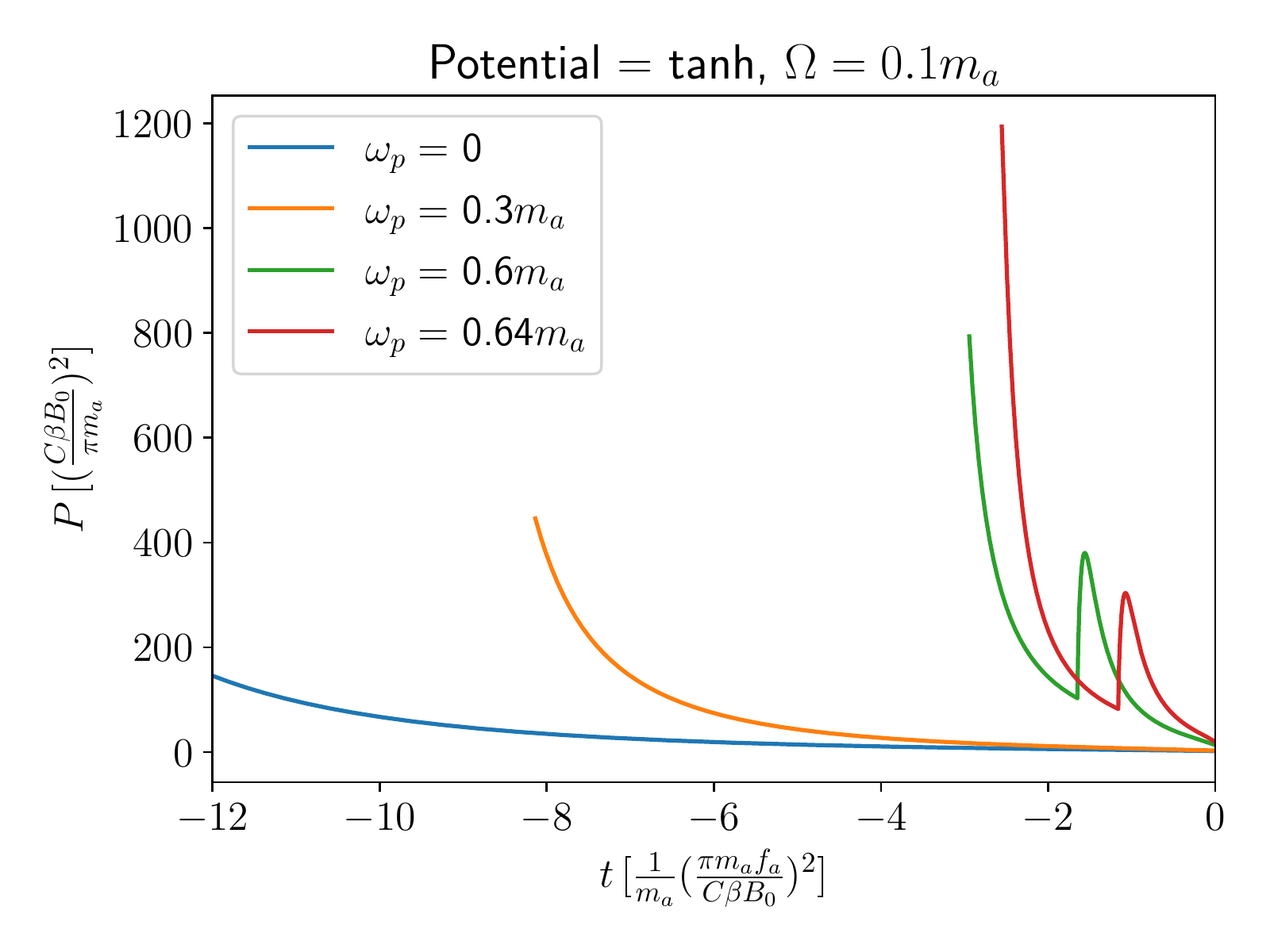}
	\end{subfigure}
	\hfill
	\begin{subfigure}{0.49\textwidth}
		\includegraphics[width=\textwidth]{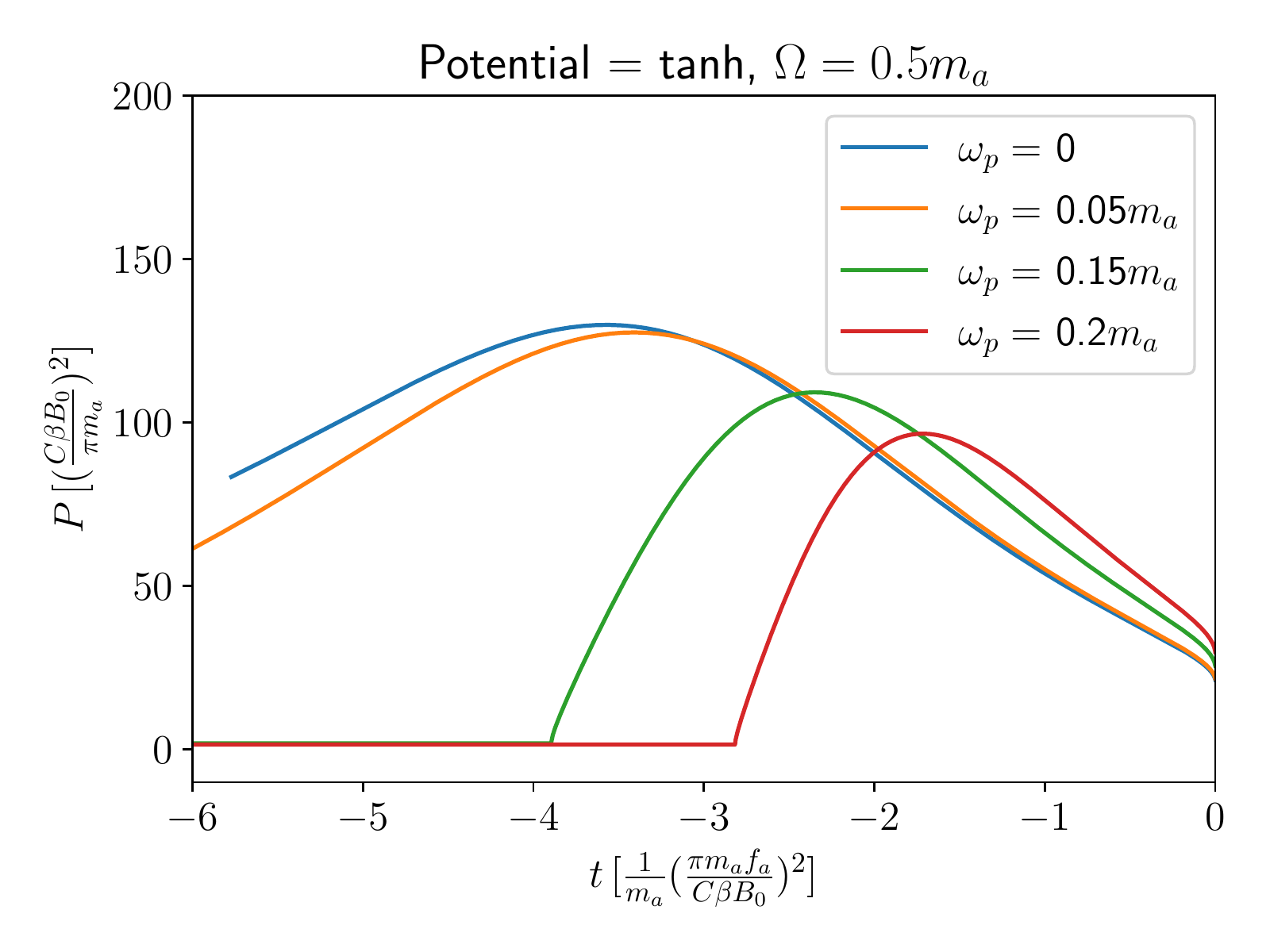}
	\end{subfigure}
	\caption{Power radiated as a function of time for the tanh-potential for a few different values of the external magnetic field frequency $\Omega$ and the plasma frequency $\op$. Again, we have $\omega(t=0) = \omega_\text{crit}$.}\label{P_of_time_tanh}
\end{figure}

\begin{figure}
	\begin{subfigure}{0.32\textwidth}
		\includegraphics[width=\textwidth]{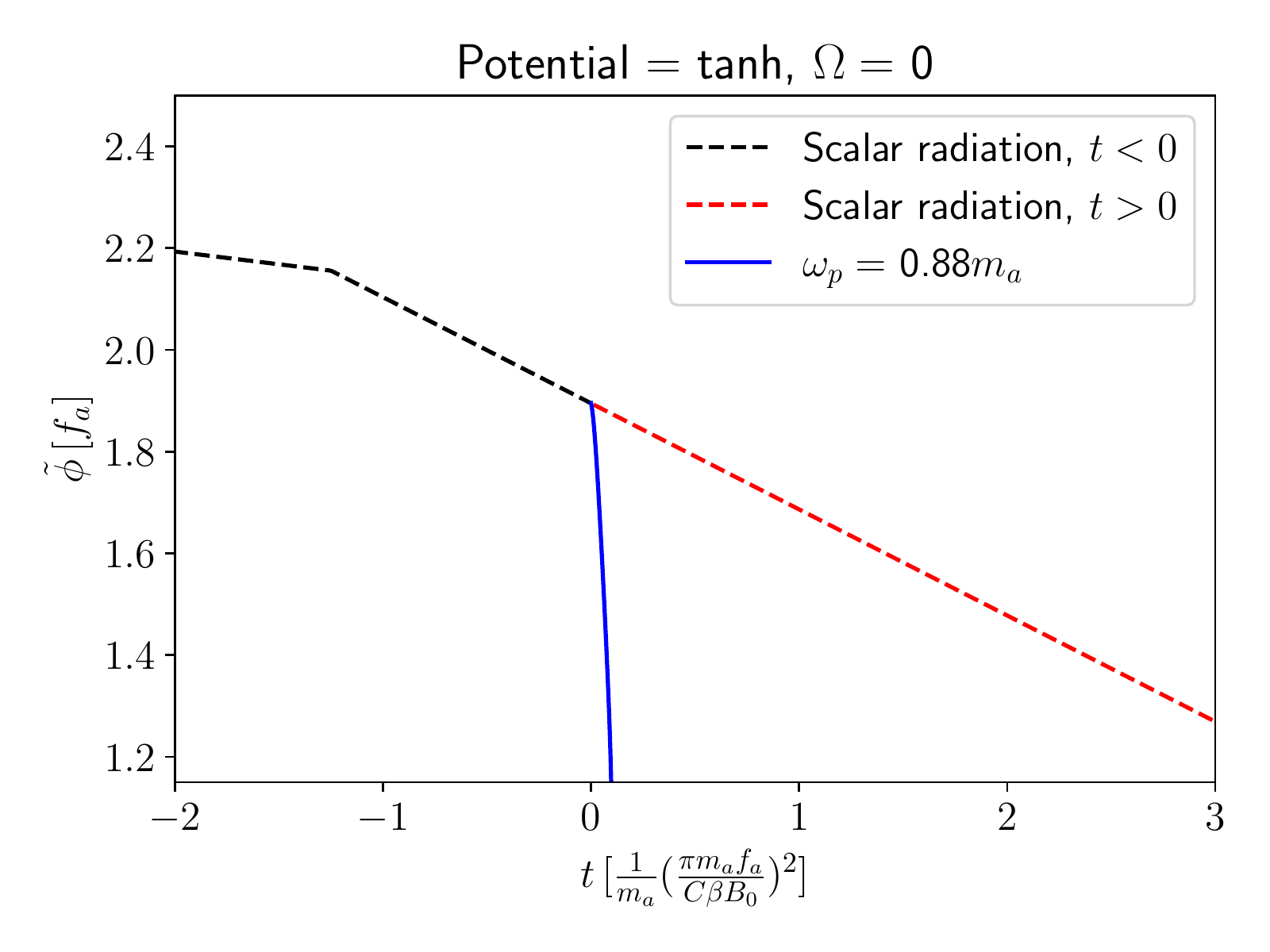}
	\end{subfigure}
	\hfill
	\begin{subfigure}{0.32\textwidth}
		\includegraphics[width=\textwidth]{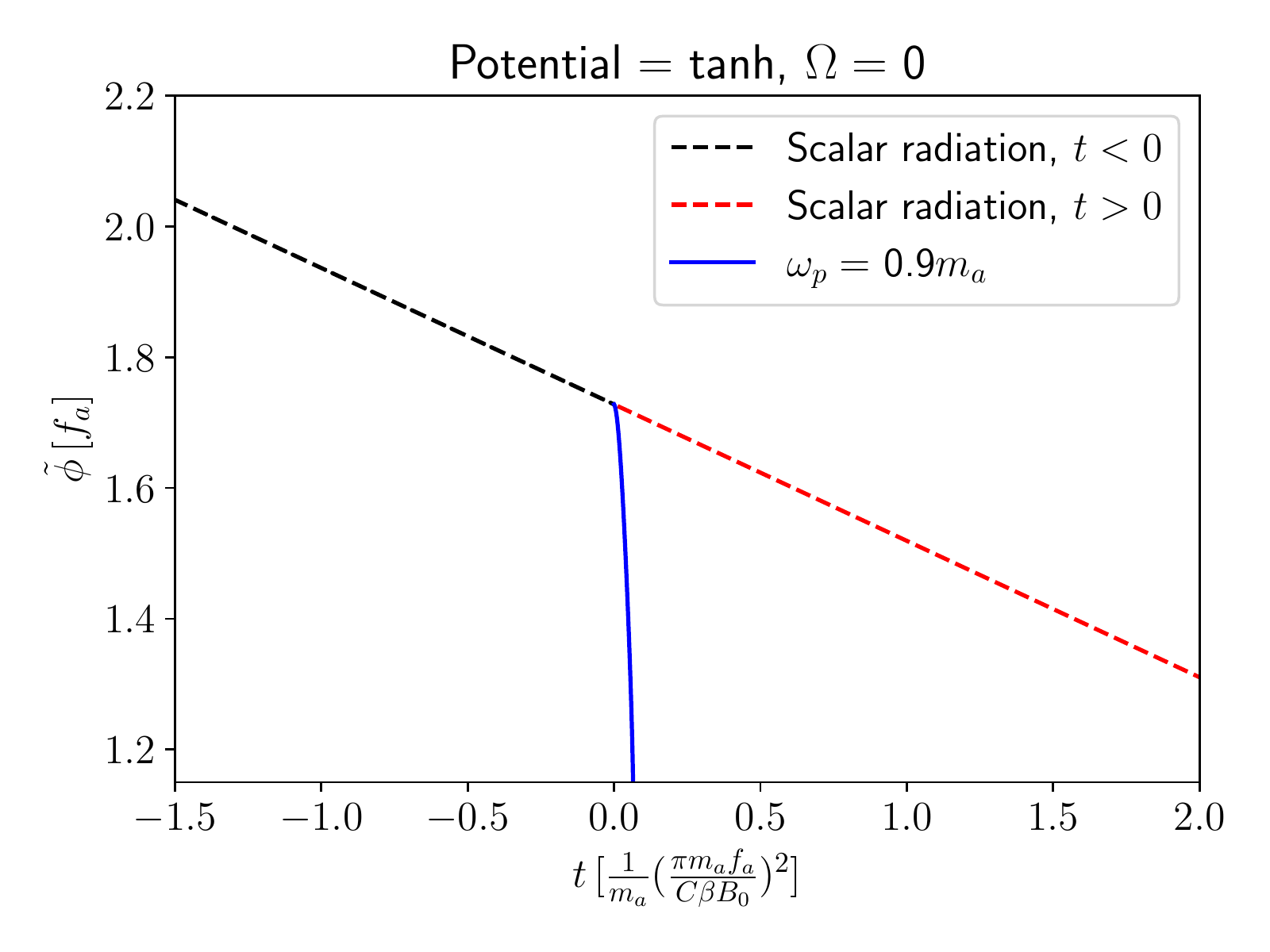}
	\end{subfigure}
	\hfill
	\begin{subfigure}{0.32\textwidth}
		\includegraphics[width=\textwidth]{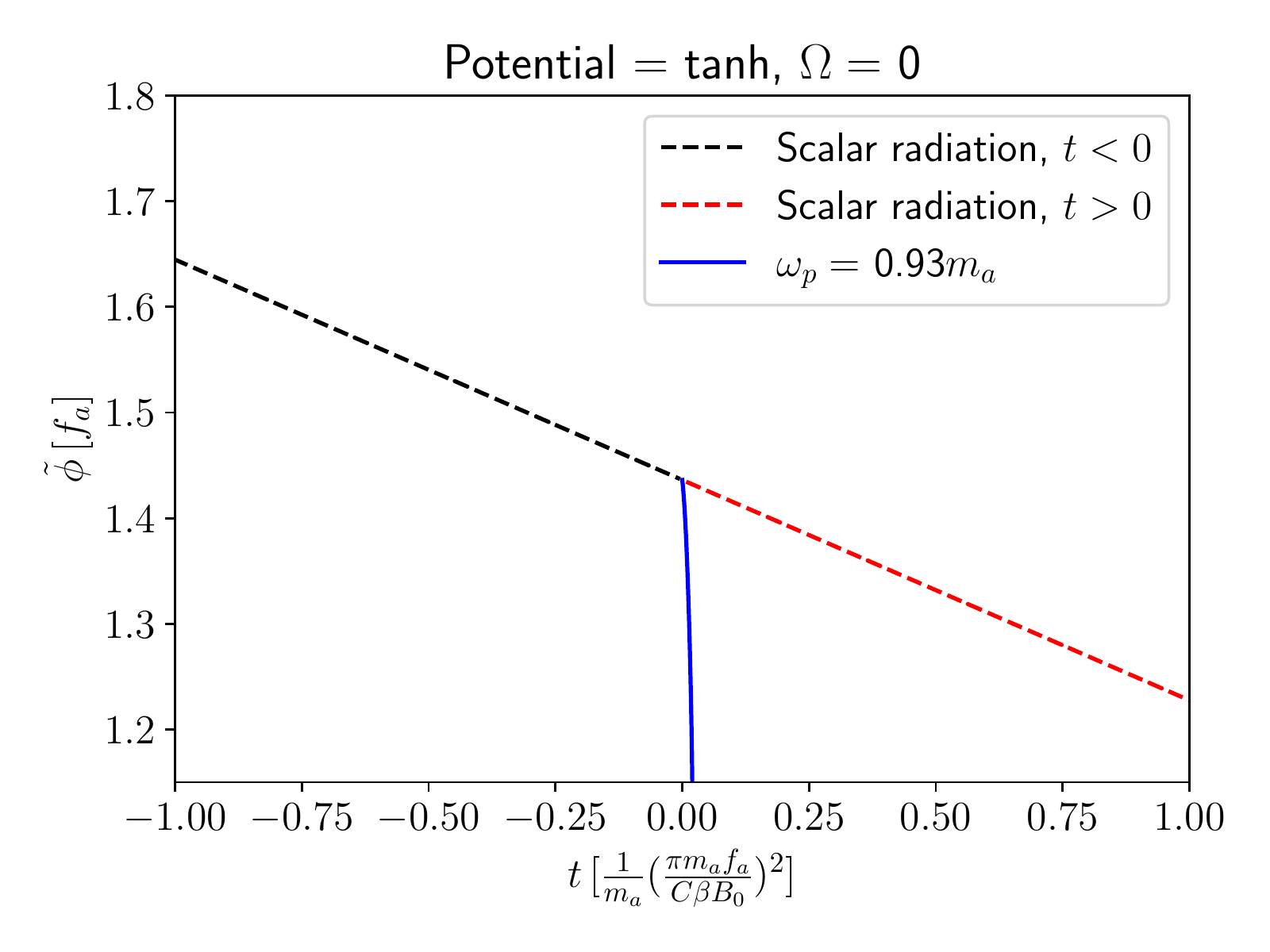}
	\end{subfigure}
	\caption{Decay of the central amplitude as it would evolve through solely scalar radiation (dashed lines) and EM radiation (solid lines), for three different values of the plasma frequency when the initial frequency satisfies $\omega(t=t_0)<\op$. We define the time $t=0$ as the time at which $\omega(t=0) = \op$ and we have chosen the constant $\Big(\frac{\pi m_a f_a}{C\beta B_0}\Big)^2=3\cdot 10^4$. The data for the scalar radiation is loosely read from figure 4 in \cite{Zhang:2020bec}. }\label{Compare_scalar_EM}
\end{figure}

\begin{figure}
	\begin{subfigure}{0.49\textwidth}
		\includegraphics[width=\textwidth]{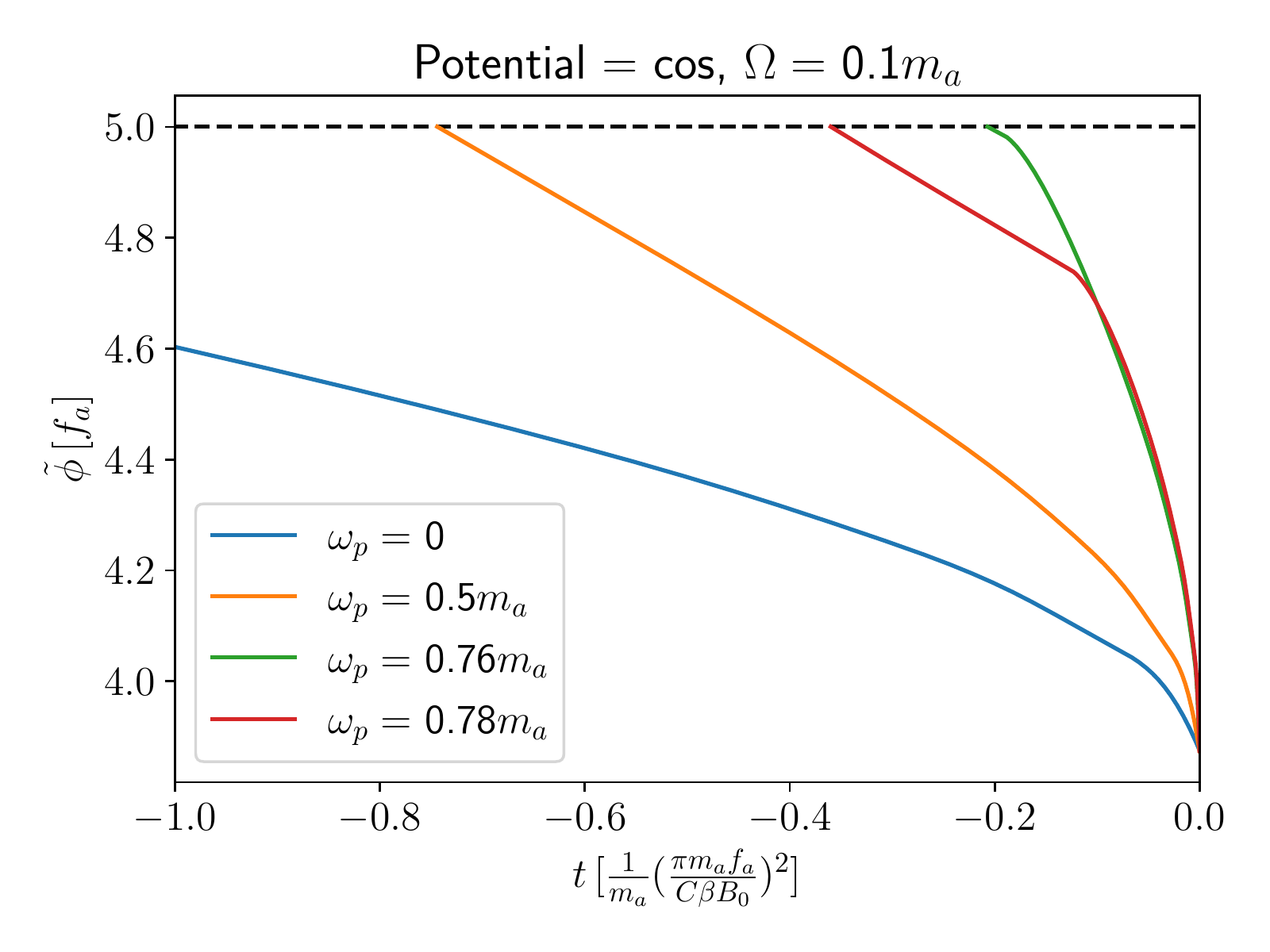}
	\end{subfigure}
	\hfill
	\begin{subfigure}{0.49\textwidth}
		\includegraphics[width=\textwidth]{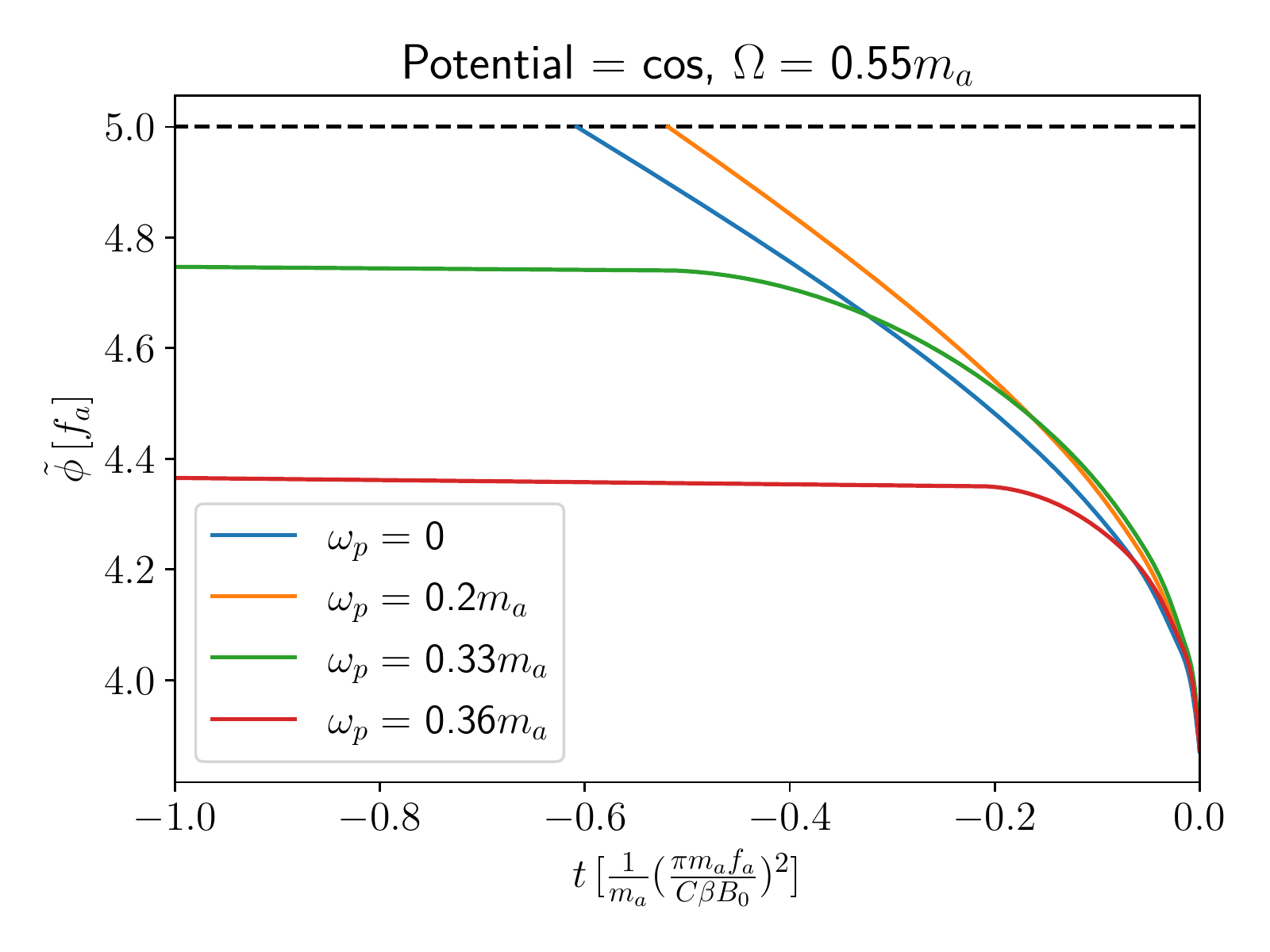}
	\end{subfigure}
	\caption{Plots show the central amplitude of the axion clump in the case of a cos-potential for a few different values of the external magnetic field frequency $\Omega$ and the plasma frequency $\op$. The dashed horizontal line represents the initial central amplitude of the clump $\phia(t_0) = 5f_a$ with $\omega(t=0) = \omega_\text{crit}$.}\label{phi0_of_time_cos}
\end{figure}

\begin{figure}
	\begin{subfigure}{0.49\textwidth}
		\includegraphics[width=\textwidth]{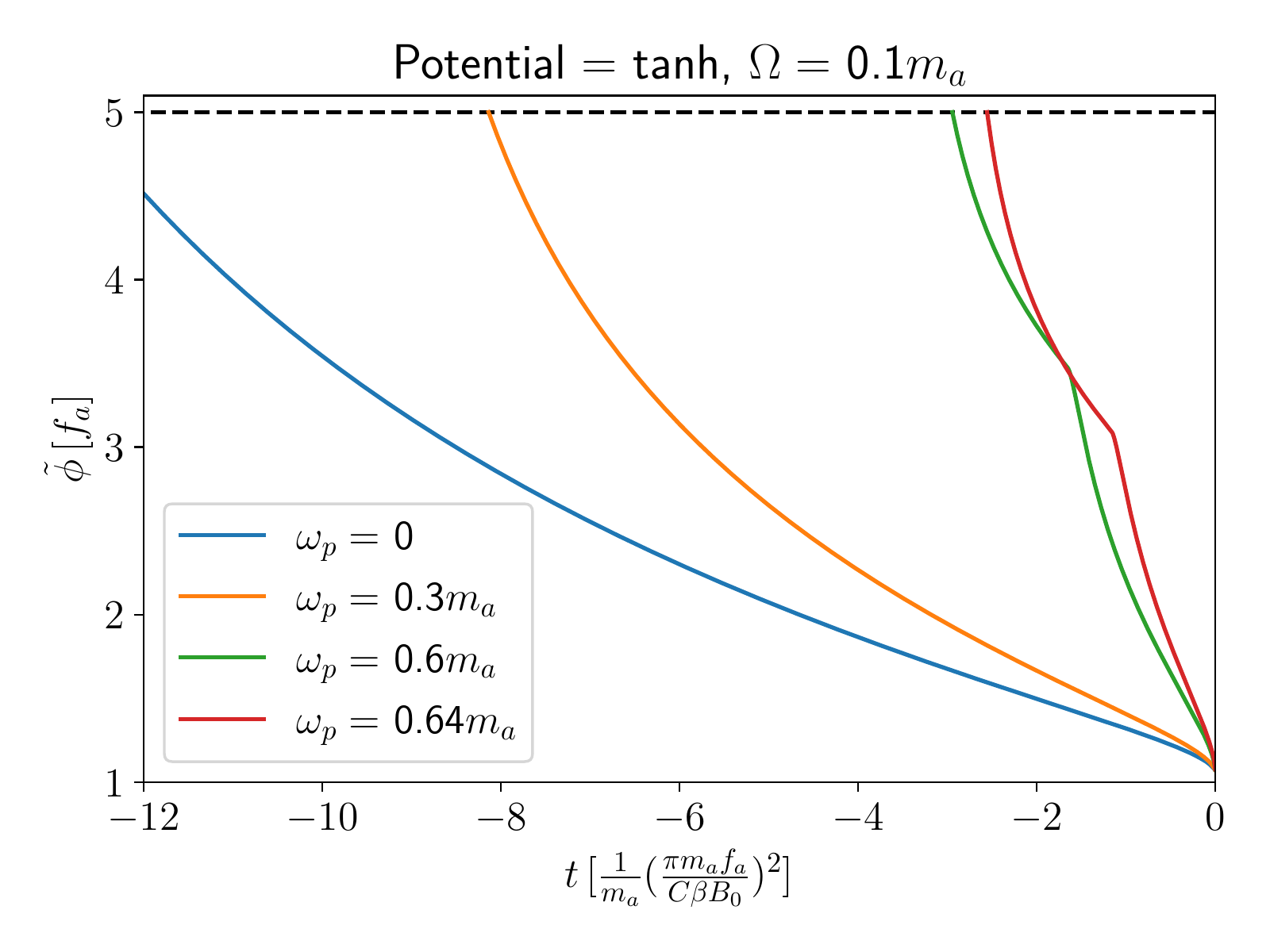}
	\end{subfigure}
	\hfill
	\begin{subfigure}{0.49\textwidth}
		\includegraphics[width=\textwidth]{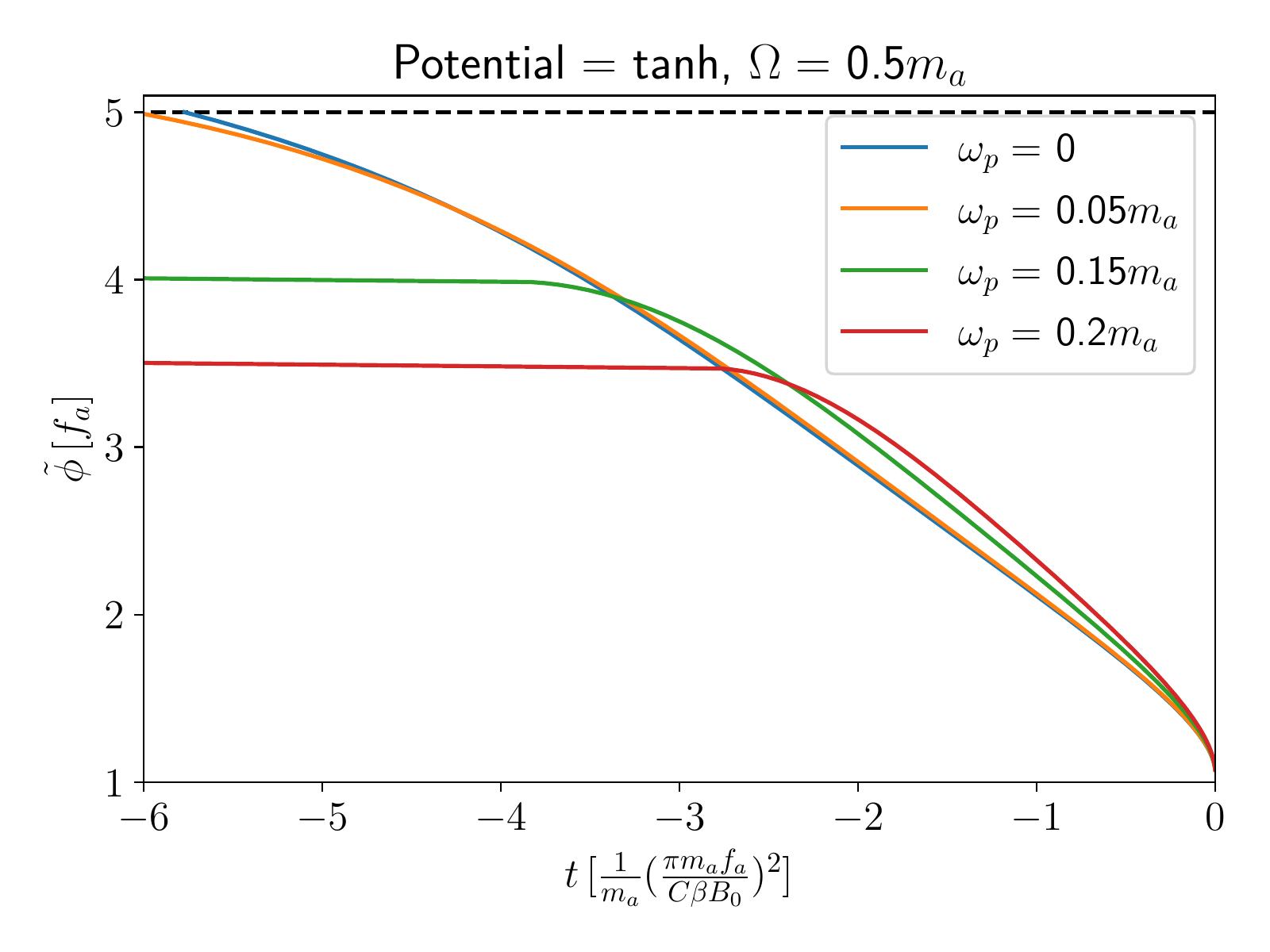}
	\end{subfigure}
	\caption{Central amplitude of the axion clump in the case of a tanh-potential for a few different values of the external magnetic field frequency $\Omega$ and the plasma frequency $\op$. The dashed horizontal line represents the frequency with the initial central amplitude of the clump $\phia(t_0)=5 f_a$ with $\omega(t=0) = \omega_\text{crit}$.}\label{phi0_of_time_tanh}
\end{figure}

We now focus our attention on $\Omega\neq 0$ and consider Fig. \ref{phi0_of_time_cos} and \ref{phi0_of_time_tanh}, former for the cos potential and latter for the tanh potential. The plasma frequencies and the external magnetic field frequency are chosen such that the clump decays shown in these figures correspond to the radiation plots in Fig. \ref{P_of_time_cos} and \ref{P_of_time_tanh}.  Of course, we initialize the clumps with $\phia=5f_a$ for both potentials to match to Fig. \ref{P_of_time_cos} and \ref{P_of_time_tanh}. The dashed line on the top of the figures indicate the $\phia=5f_a$ line. The axis at $t=0$ corresponds to the instant in time where the clumps reach critical frequency and therefore become unstable. The scale on the horizontal time axis has been chosen to accurately depict the resonances at the cost of not showing the entire lifespan of the slowest decaying clumps. 
In Fig \ref{phi0_of_time_cos} for $\Omega=0.1m_a$, the blue curve with a plasma frequency of $\omega_P=0$ decays over a timescale $\tau_\text{EM}\sim 2\cdot\frac{1}{m_a}\Big(\frac{\pi m_a f_a}{C\beta B_0}\Big)^2$ which is consistent with the estimate  \eqref{tau}, but about an order of magnitude slower than the green and red curves with $\op=0.76m_a$ and $\op=0.78m_a$, respectively. This is why, in the left panel figure we don't see the blue curve touch the dashed horizontal line on the top. On the left panel, the end points of the red, green and orange curve for $t<0$ for $\Omega=0.1 m_a$, touching the $\phia=5f_a$ line, are the points where the corresponding clumps are initialized. Similarly, for the right panel, the blue and the orange curves touch the $\phia=5f_a$ line where the clumps are initialized. The corresponding initialization points for the red and green curves are outside of the range of the axis shown, i.e. further back in time.  
The sudden increase in radiation for the red and green curves in the left panel in Fig. \ref{P_of_time_cos} translates to the sudden increase in decay rate of the clumps seen in the left panel of Fig. \ref{phi0_of_time_cos}. This effect is significantly more pronounced for the red and green curves for $\Omega=0.55 m_a$ on the right panel. Note that, for the red and green curves with $\op=0.33 m_a, 0.36 m_a$, the clumps take a long time to decay from a values of $5f_a$ to about $4.8 f_a$ and $4.4 f_a$ in Fig. \ref{P_of_time_cos}. This EM decay timescale is more than an order of magnitude larger the estimate in Eq. \ref{tau}. However, as soon as the clumps reach resonance, the decay rate enhances, forcing the clumps to decay significantly over a timescale that is shorter than the estimate in Eq. \ref{tau} by an order of magnitude. 

\begin{figure}
	\begin{subfigure}{0.49\textwidth}
		\includegraphics[width=\textwidth]{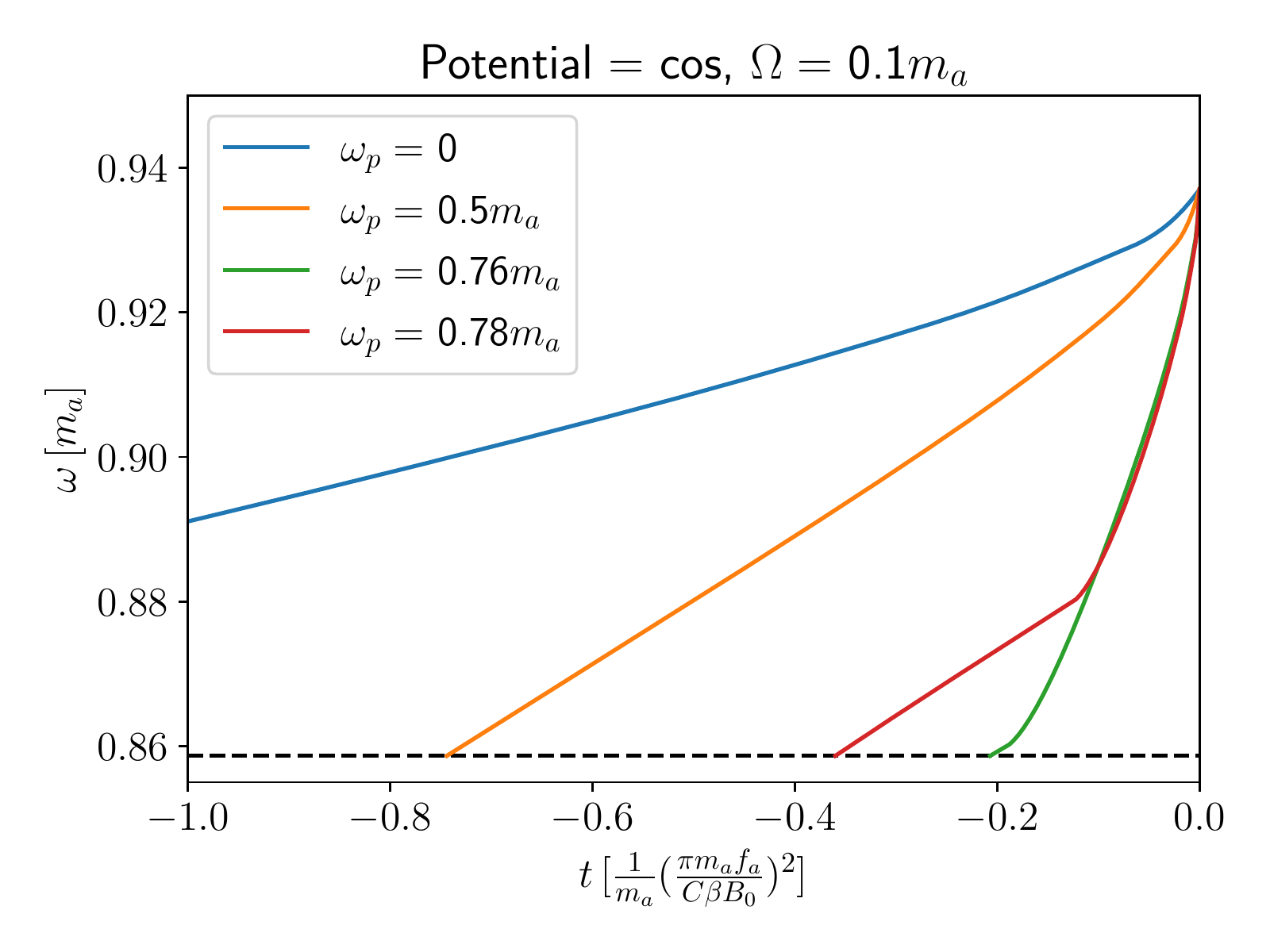}
	\end{subfigure}
	\hfill
	\begin{subfigure}{0.49\textwidth}
		\includegraphics[width=\textwidth]{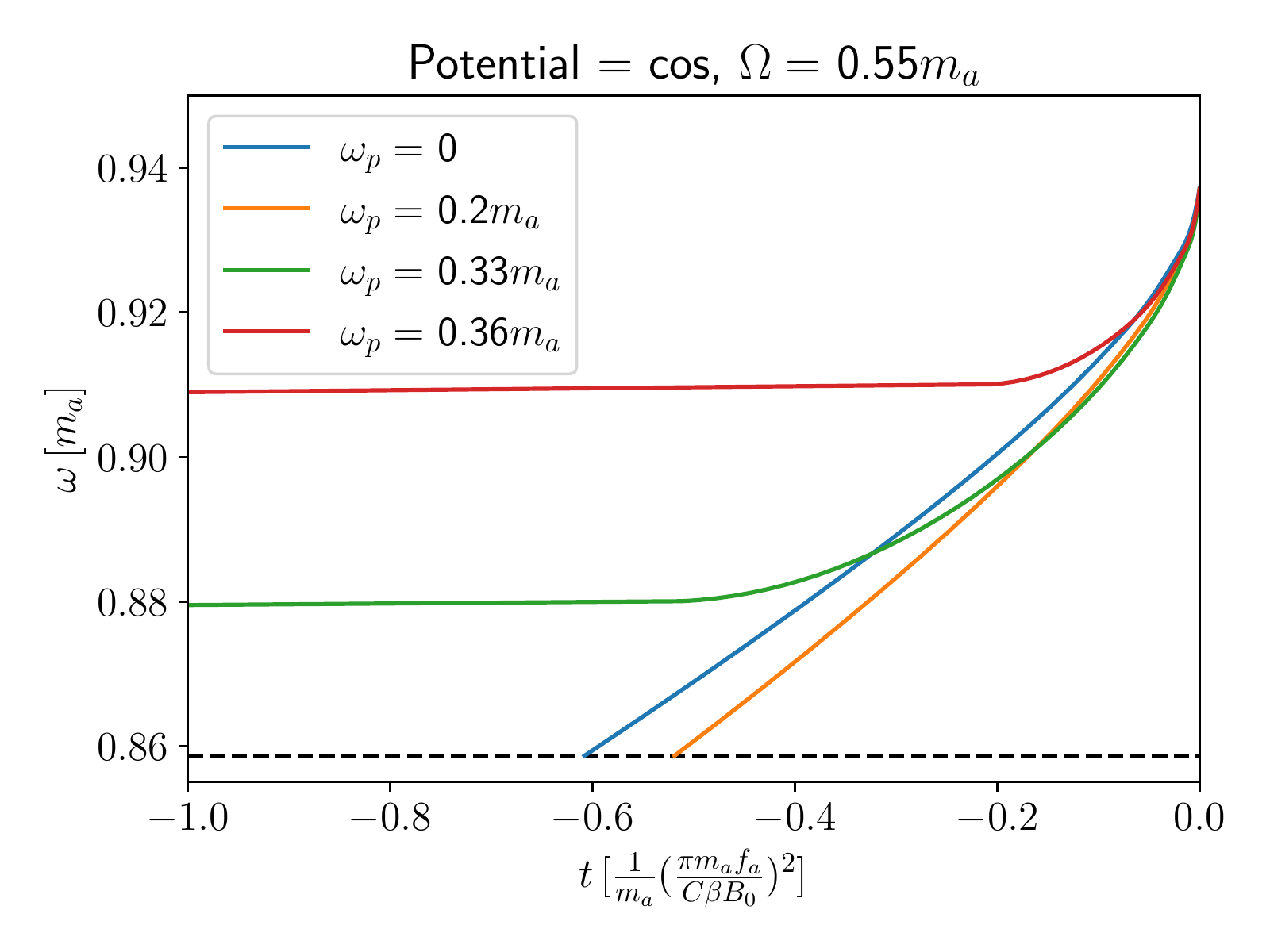}
	\end{subfigure}
	\caption{Clump frequency $\omega$ as a function of time for the cos-potential for a few different values of the external magnetic field frequency $\Omega$ and the plasma frequency $\op$. The dashed horizontal line shows the frequency corresponding to the initial central amplitude of the clump $\phia(t_0) = 5f_a$. As before $\omega(t=0) = \omega_\text{crit}$}\label{omega_of_time_cos}
\end{figure}

\begin{figure}
	\begin{subfigure}{0.49\textwidth}
		\includegraphics[width=\textwidth]{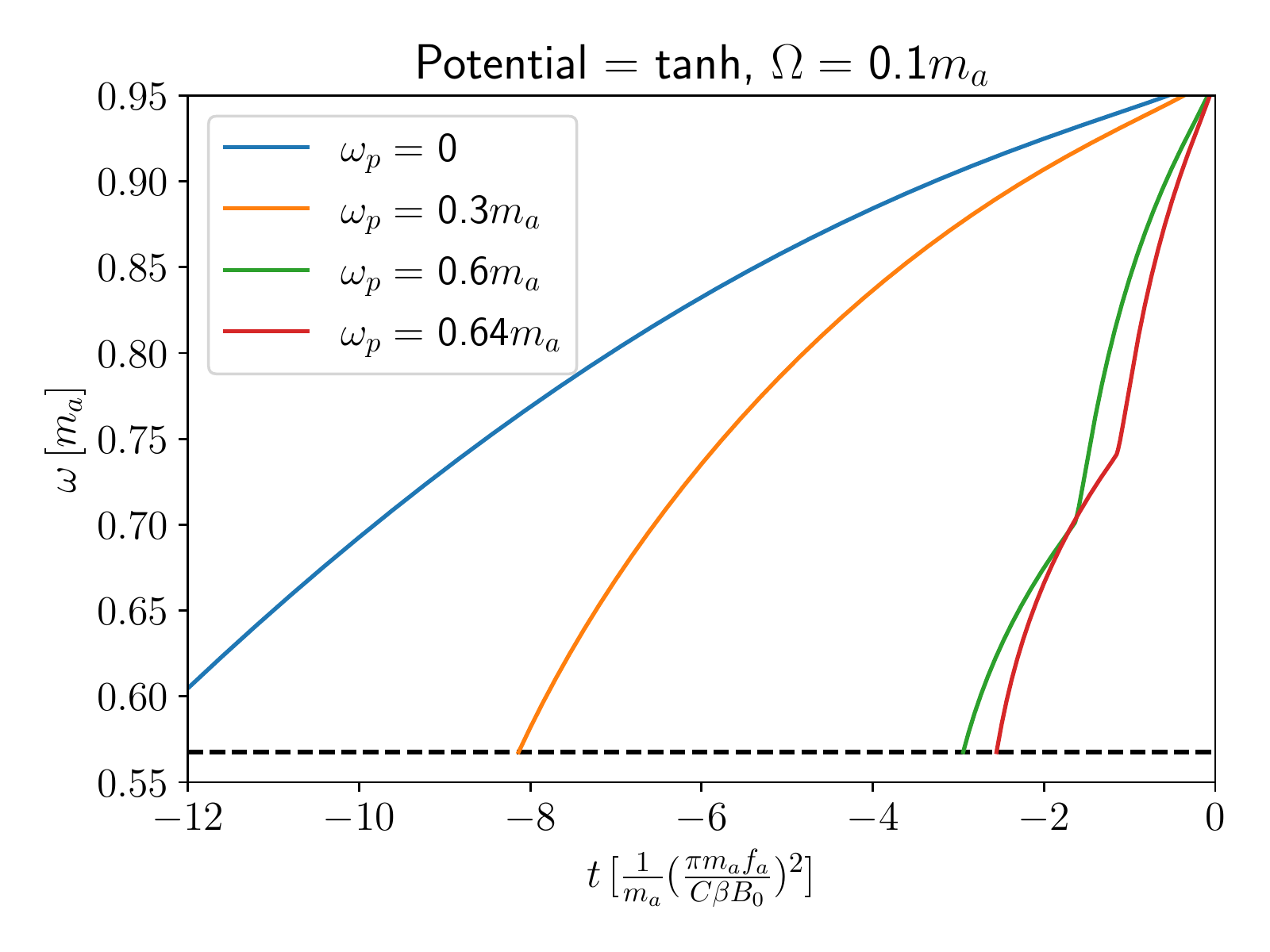}
	\end{subfigure}
	\hfill
	\begin{subfigure}{0.49\textwidth}
		\includegraphics[width=\textwidth]{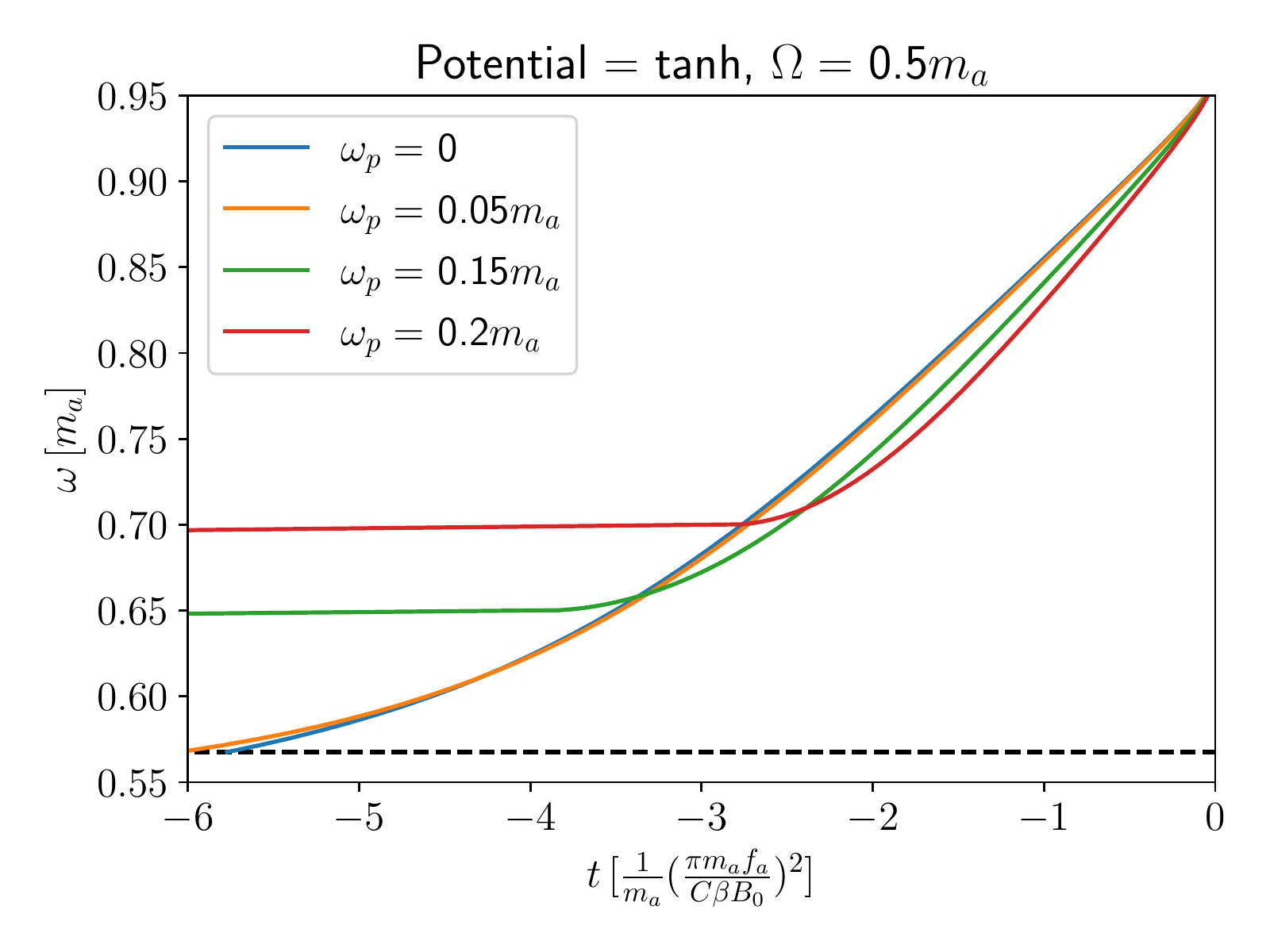}
	\end{subfigure}
	\caption{Clump frequency $\omega$ as a function of time for the tanh-potential for a few different values of the external magnetic field frequency $\Omega$ and the plasma frequency $\op$. The dashed horizontal line represents the frequency corresponding to the initial clump amplitude $\phia(t_0) = 5f_a$ and $\omega(t=0) = \omega_\text{crit}$.}\label{omega_of_time_tanh}
\end{figure}

\begin{figure}
	\begin{subfigure}{0.49\textwidth}
		\includegraphics[width=\textwidth]{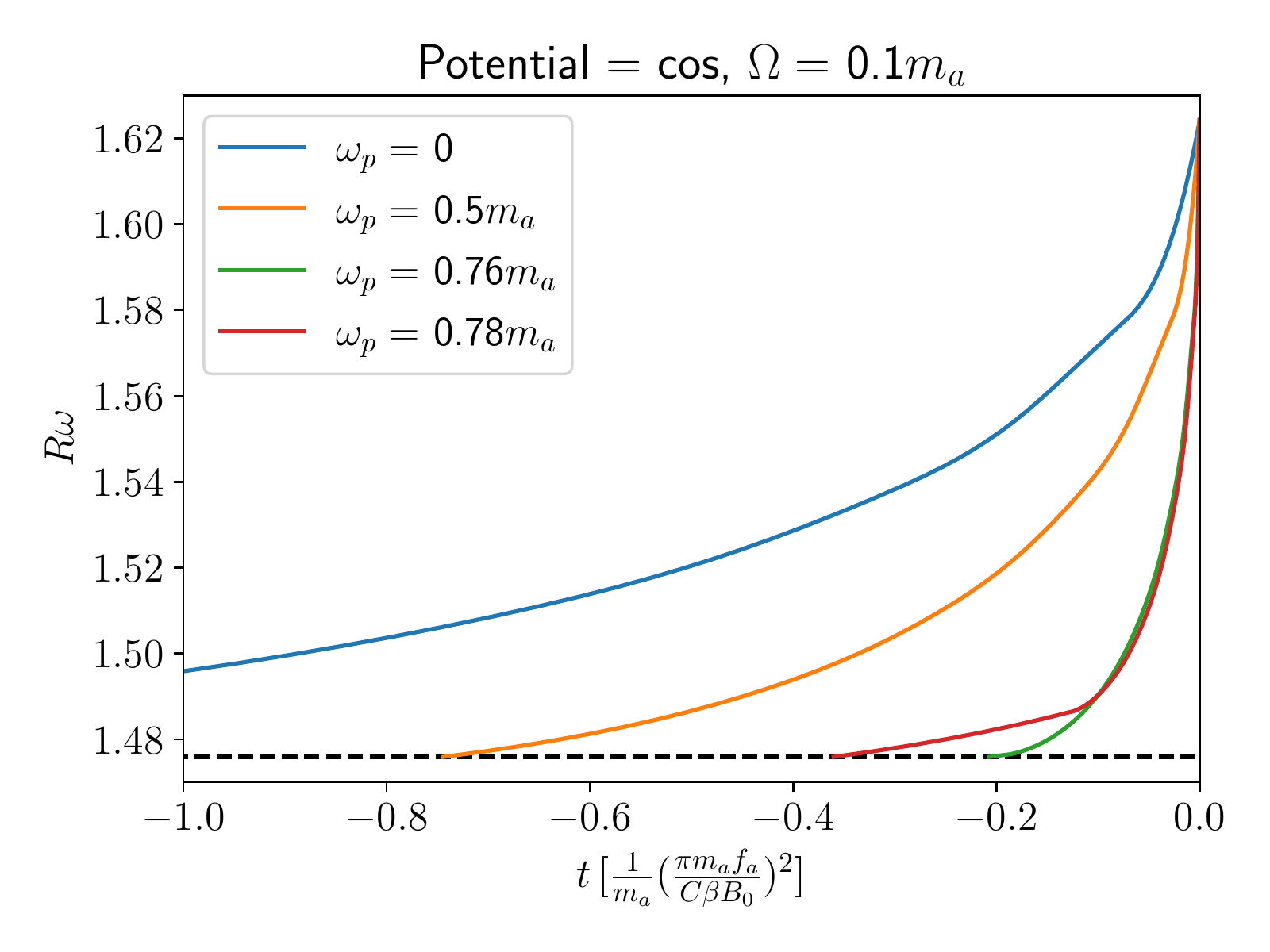}
	\end{subfigure}
	\hfill
	\begin{subfigure}{0.49\textwidth}
		\includegraphics[width=\textwidth]{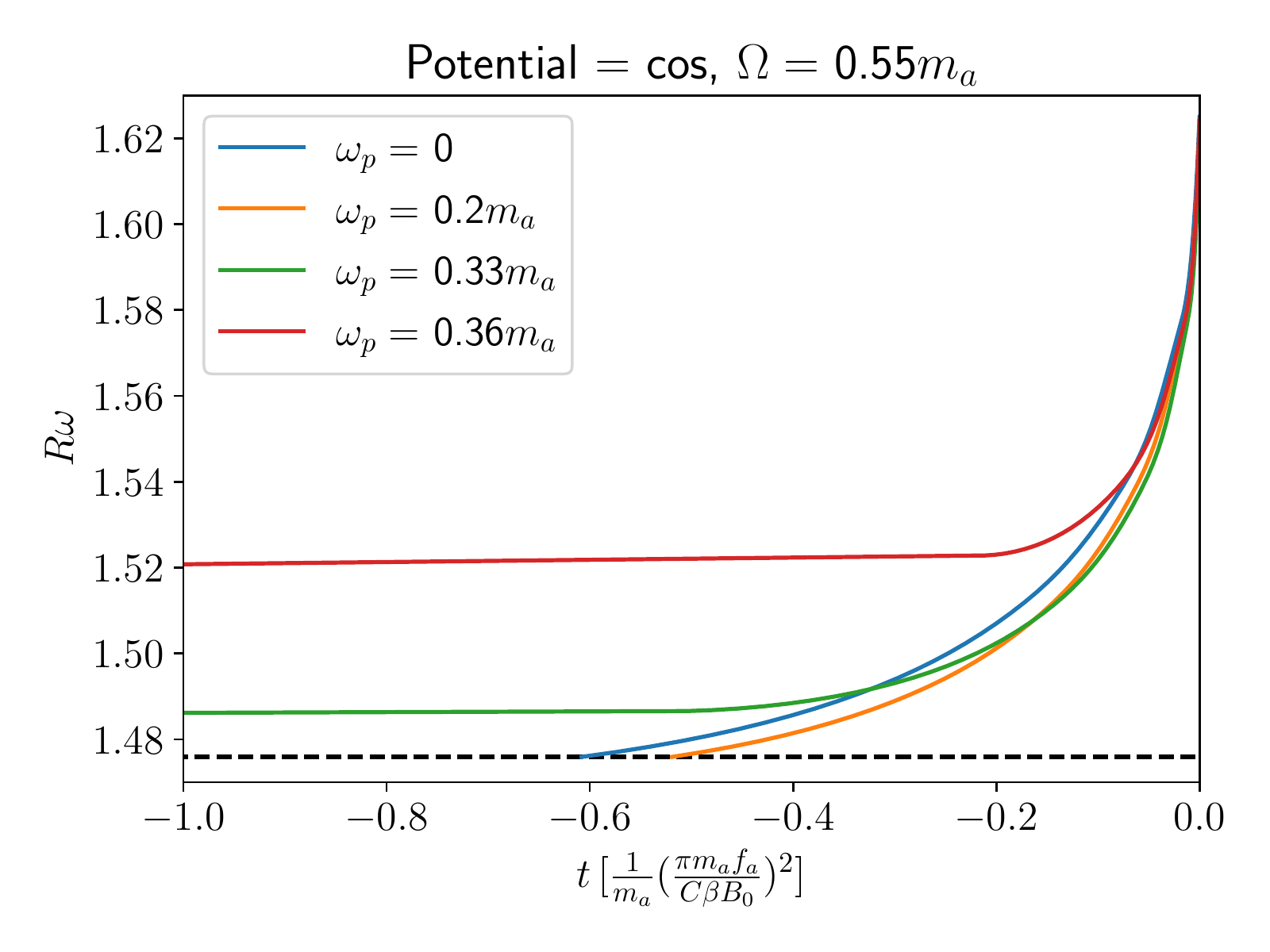}
	\end{subfigure}
	\caption{Figures shows the product $\omega R$ as a function of time for the cos-potential for a few different values of the external magnetic field frequency $\Omega$ and the plasma frequency $\op$. The dashed horizontal line shows the initial value for the same quantity. Time axis is chosen such that $\omega(t=0)=\omega_\text{crit}$.}\label{R_of_time_cos}
\end{figure}

\begin{figure}
	\begin{subfigure}{0.49\textwidth}
		\includegraphics[width=\textwidth]{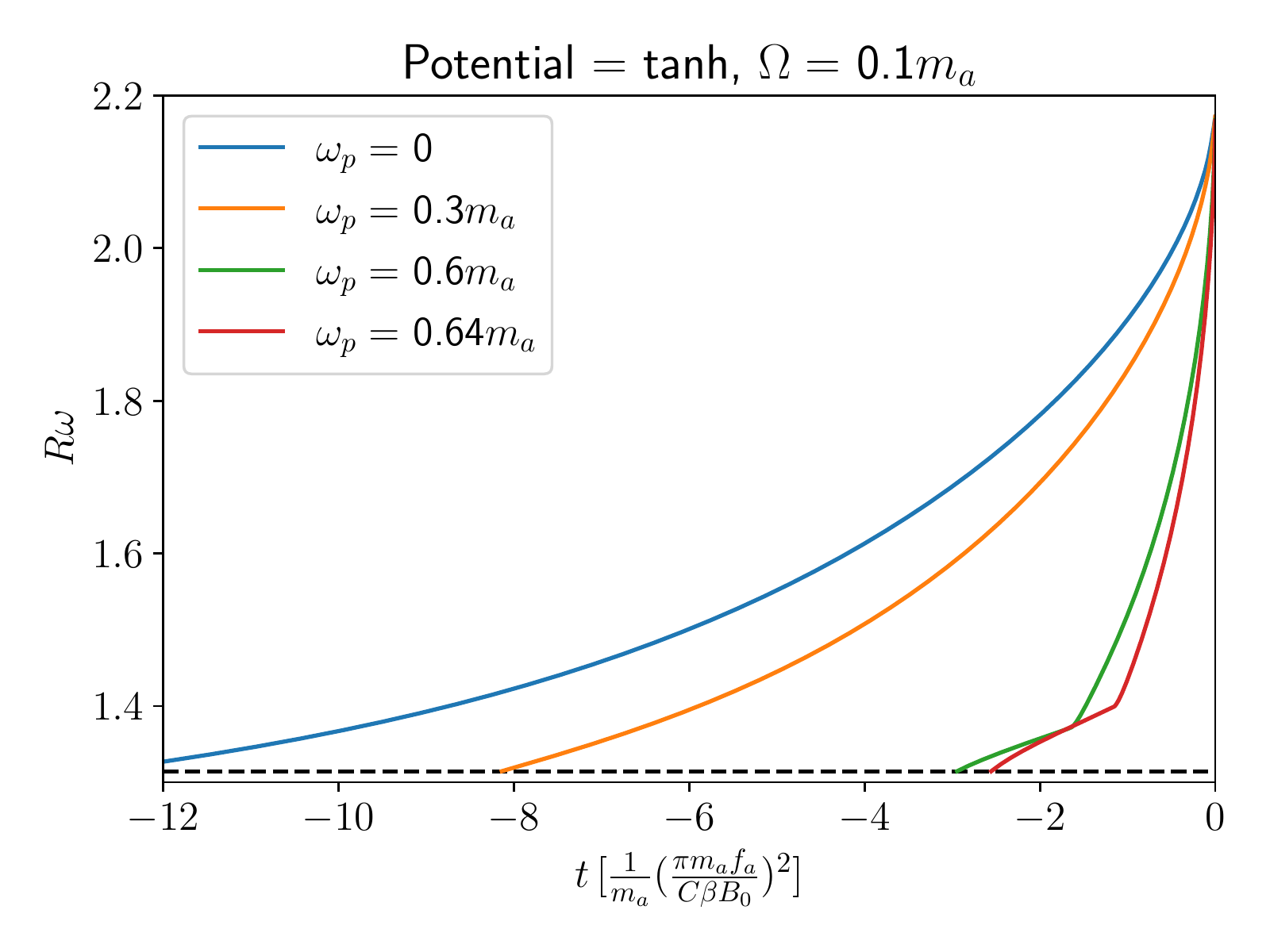}
	\end{subfigure}
	\hfill
	\begin{subfigure}{0.49\textwidth}
		\includegraphics[width=\textwidth]{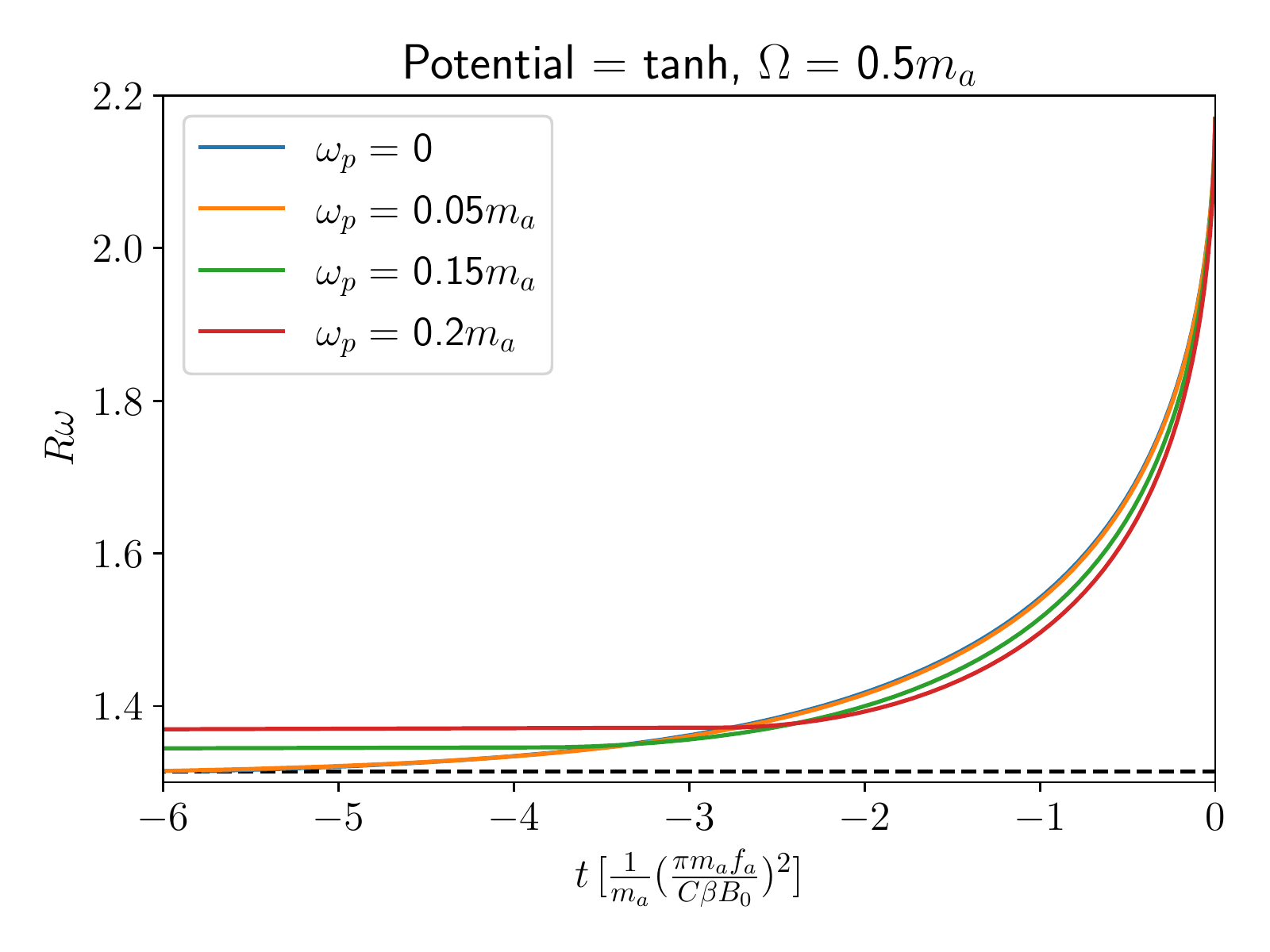}
	\end{subfigure}
	\caption{The product $R\omega$ as a function of time for the tanh-potential for a few different values of the external magnetic field frequency $\Omega$ and the plasma frequency $\op$. The dashed horizontal line shows the corresponding initial value, with $\omega(t=0) = \omega_\text{crit}$.}\label{R_of_time_tanh}
\end{figure}

We can see a very similar behavior for the tanh potential in Fig. \ref{phi0_of_time_tanh}. Figures \ref{omega_of_time_cos} and \ref{omega_of_time_tanh} show the corresponding time evolution for the frequency of the clump. The dashed horizontal lines indicate the value of $\omega$ at the instant of initialization. Figures \ref{R_of_time_cos} and \ref{R_of_time_tanh} show the behavior of $R\omega$ as a function of time. From these figures we can see that a sudden increase in radiation seen in Fig. \ref{P_of_time_cos} and \ref{P_of_time_tanh} affects the time evolution of the frequency and the size of the clump quite significantly.

\subsection{Decay time-scales and regime of validity of our approach}\label{decay_timescale}
We now take a minute to look closer at the timescale of the decay for the axion clumps considered. This will also help us establish the regime of validity of the calculational strategy we employed for taking into account back-reaction. 
Note that for our calculations to be valid and our approximations to hold we need a few different conditions to be satisfied. First of all we need to satisfy the weak the coupling condition such that $\frac{C\beta}{\pi}<1$. Similarly, we will need the electromagnetic decay timescale to satisfy $\frac{2\pi}{\omega}<\tau_{EM}<\tau_{\text{scalar}}$ where $\tau_{\text{scalar}}$ is the decay timescale for scalar radiation. The conditions $\frac{2\pi}{\omega}<\tau_{EM}$ and $\frac{C\beta}{\pi}<1$ are easily satisfied. We need to verify under what condition $\tau_{EM}<\tau_{\text{scalar}}$. As seen from the amplitude decay plots of Fig. \ref{phi0_of_time_cos} and \ref{phi0_of_time_tanh}, the decay timescales are typically 
given by the estimate in Eq. \ref{tau}. In fact, when clumps start resonantly radiating, the corresponding decay timescale is even smaller. Our goal is to compare this timescale with the scalar radiation timescale.

For the cos potential the decay timescale from scalar radiation was found to be about $10^3 m_a^{-1}$ and from tanh potential it's about $10^6 m_a^{-1}$ \cite{Zhang:2020bec,Piette_1998}. Therefore we see that as long as our parameters are such that $\Big(\frac{\pi m_a f_a}{C\beta B_0}\Big)^2<10^3$ for the cos potential and 
$\Big(\frac{\pi m_a f_a}{C\beta B_0}\Big)^2<10^6$ for the tanh potential, our calculations for back-reaction remain valid. In order to understand what this condition implies for the QCD axion as an example, we set $\beta=1/137$ with $C=1$. Setting $m_af_a\sim \Lambda_{\text{QCD}}^2$, we need $B_0>13 \Lambda_{\text{QCD}}^2$ and $B_0>0.43 \Lambda_{\text{QCD}}^2$ for the cos and the tanh potential. It is difficult to find such high magnetic fields in nature. However, if we consider axion-like particles, then $\beta$ can be taken larger than $1/137$ while also maintaining $C\beta/\pi<1$ and $m_a f_a\neq \Lambda_{\text{QCD}}^2$. Therefore it is easier to satisfy the constraints with moderate magnetic fields in order to stay within the regime of validity. Note that, the electromagnetic decay timescale is set by the clump frequency, its central amplitude, the EM coupling as well as the strength of the external magnetic field. The details of the axion self interaction can only play a role in determining the validity of our calculation. It does not directly affect the electromagnetic decay timescale.

\section{Conclusion}\label{conclusion}
In this paper our goal was to describe electromagnetic radiation from axion clumps in a magnetic field while taking into account back-reaction of the clump. Our approach is semi-analytic and perturbative, therefore complements previous efforts \cite{Amin:2021tnq,Sen:2021mhf} which employ a complete numerical treatment of axion-photon equations. Such numerical implementation may in some cases be computationally expensive. In particular, when the electromagnetic coupling of the axions is weak, a complete numerical analysis may be unnecessary. As we showed in this paper, the energy taken away in the form of electromagnetic radiation reduces the total energy of the axion clump in such a way that at every instant in time the clump configuration approximately satisfies the sourceless ($C\beta=0$) axion equation of motion. In other words, the axion configuration is primarily set by the axion self-interaction while maintaining energy conservation.  


As noted in \cite{Zhang:2020bec}, the axion EOM gives rise to clump solutions where the central clump amplitude, the axion frequency and the clump size are related to each other and the total particle number in the clump. In fact, one finds that the clump frequency increases as a function of the central amplitude. This coupled with the observation in \cite{Amin:2021tnq,Sen:2021mhf} that electromagnetic radiation from an axion clump can undergo resonant enhancement when its frequency matches with the plasma frequency, served as one of the motivations for our analysis. We show that a clump that doesn't satisfy resonant condition at a certain instant in time can radiate slowly, thereby altering its frequency with time in such a way that the resonant condition is satisfied at a later time leading to a rapid increase in radiated power. On the other hand, the opposite can be true too. A clump that radiates efficiently at some initial time can move out of resonance leading to a reduction in radiation. Neither effect can be seen unless one takes into account back-reaction of axion clumps. 

Note that in most of the paper we don't take into account scalar waves radiation explicitly. We do however take scalar radiation into account when EM radiation is absent at early times as in the case of a clump with a static background magnetic field with $\omega<\omega_P$. In all other cases we work in a regime where scalar wave emission is suppressed compared to the EM radiation and clumps primarily decay through electromagnetic emission. Our perturbative semi-analytic analysis is reliant on electromagnetic decay timescale being smaller than the scalar wave decay scale. We identified the parameter space for which this holds. It is important to remember that whether a clump moves in or out of resonance as a function of time is not constrained by such consideration. These considerations are only important for us to be able to do a perturbative analytic calculation. Therefore to describe a scenario where both scalar radiation and EM radiation are significant, one needs to take into account both. 
Such an analysis is beyond the scope of this paper and will be pursued in future work. 

The regime of parameter space inaccessible to us also includes very large dilute clumps with $R\gg m_a^{-1}$. This is because in this regime gravity becomes non-negligible compared to axion self-interactions. Since we throughout have ignored gravity in our calculations, this is beyond the scope of this paper and is deferred for future work. 
\\ \\
\noindent\textbf{Acknowledgments}
\\ \\
This work was supported by the Department of Energy Nuclear Physics Quantum Horizons program through the Early Career Award DE-SC0021892.

\bibliographystyle{JHEP}
\bibliography{ref}

\providecommand{\href}[2]{#2}\begingroup\raggedright\begin{thebibliography}{10}

\bibitem{PhysRevLett.38.1440}
R.D.~Peccei and H.R.~Quinn, \emph{$\mathrm{CP}$ conservation in the presence of
  pseudoparticles},
  \href{https://doi.org/10.1103/PhysRevLett.38.1440}{\emph{Phys. Rev. Lett.}
  {\bfseries 38} (1977) 1440}.

\bibitem{PhysRevLett.40.223}
S.~Weinberg, \emph{A new light boson?},
  \href{https://doi.org/10.1103/PhysRevLett.40.223}{\emph{Phys. Rev. Lett.}
  {\bfseries 40} (1978) 223}.

\bibitem{DINE1983137}
M.~Dine and W.~Fischler, \emph{The not-so-harmless axion},
  \href{https://doi.org/https://doi.org/10.1016/0370-2693(83)90639-1}{\emph{Physics
  Letters B} {\bfseries 120} (1983) 137}.

\bibitem{Preskill:1982cy}
J.~Preskill, M.B.~Wise and F.~Wilczek, \emph{{Cosmology of the Invisible
  Axion}}, \href{https://doi.org/10.1016/0370-2693(83)90637-8}{\emph{Phys.
  Lett. B} {\bfseries 120} (1983) 127}.

\bibitem{Abbott:1982af}
L.F.~Abbott and P.~Sikivie, \emph{{A Cosmological Bound on the Invisible
  Axion}}, \href{https://doi.org/10.1016/0370-2693(83)90638-X}{\emph{Phys.
  Lett. B} {\bfseries 120} (1983) 133}.

\bibitem{Kim:1986ax}
J.E.~Kim, \emph{{Light Pseudoscalars, Particle Physics and Cosmology}},
  \href{https://doi.org/10.1016/0370-1573(87)90017-2}{\emph{Phys. Rept.}
  {\bfseries 150} (1987) 1}.

\bibitem{Cheng:1987gp}
H.-Y.~Cheng, \emph{{The Strong CP Problem Revisited}},
  \href{https://doi.org/10.1016/0370-1573(88)90135-4}{\emph{Phys. Rept.}
  {\bfseries 158} (1988) 1}.

\bibitem{Raffelt:1990yz}
G.G.~Raffelt, \emph{{Astrophysical methods to constrain axions and other novel
  particle phenomena}},
  \href{https://doi.org/10.1016/0370-1573(90)90054-6}{\emph{Phys. Rept.}
  {\bfseries 198} (1990) 1}.

\bibitem{Duffy:2009ig}
L.D.~Duffy and K.~van Bibber, \emph{{Axions as Dark Matter Particles}},
  \href{https://doi.org/10.1088/1367-2630/11/10/105008}{\emph{New J. Phys.}
  {\bfseries 11} (2009) 105008}
  [\href{https://arxiv.org/abs/0904.3346}{{\ttfamily 0904.3346}}].

\bibitem{Berezhiani1991}
Z.G.~Berezhiani, \emph{{On the possibility of a solution to the strong CP
  problem without axion in a SU(3)-M family symmetry model}},
  \href{https://doi.org/10.1142/S0217732391002864}{\emph{Mod. Phys. Lett. A}
  {\bfseries 6} (1991) 2437}.

\bibitem{1992SvJNP..55.1063B}
Z.G.~{Berezhiani}, A.S.~{Sakharov} and M.Y.~{Khlopov}, \emph{{Primordial
  background of cosmological axions.}}, {\emph{Soviet Journal of Nuclear
  Physics} {\bfseries 55} (1992) 1063}.

\bibitem{1994PAN....57..485S}
A.S.~{Sakharov} and M.Y.~{Khlopov}, \emph{{The nonhomogeneity problem for the
  primordial axion field}}, {\emph{Physics of Atomic Nuclei} {\bfseries 57}
  (1994) 485}.

\bibitem{1996PAN....59.1005S}
A.S.~{Sakharov}, D.D.~{Sokoloff} and M.Y.~{Khlopov}, \emph{{Large-scale
  modulation of the distribution of coherent oscillations of a primordial axion
  field in the universe}}, {\emph{Physics of Atomic Nuclei} {\bfseries 59}
  (1996) 1005}.

\bibitem{KHLOPOV1999105}
M.~Khlopov, A.~Sakharov and D.~Sokoloff, \emph{The nonlinear modulation of the
  density distribution in standard axionic cdm and its cosmological impact},
  \href{https://doi.org/https://doi.org/10.1016/S0920-5632(98)00511-8}{\emph{Nuclear
  Physics B - Proceedings Supplements} {\bfseries 72} (1999) 105}.

\bibitem{PhysRevD.99.064049}
S.D.~Odintsov and V.K.~Oikonomou, \emph{$f(r)$ gravity inflation with
  string-corrected axion dark matter},
  \href{https://doi.org/10.1103/PhysRevD.99.064049}{\emph{Phys. Rev. D}
  {\bfseries 99} (2019) 064049}.

\bibitem{PhysRevD.99.104070}
S.D.~Odintsov and V.K.~Oikonomou, \emph{Unification of inflation with dark
  energy in $f(r)$ gravity and axion dark matter},
  \href{https://doi.org/10.1103/PhysRevD.99.104070}{\emph{Phys. Rev. D}
  {\bfseries 99} (2019) 104070}.

\bibitem{Odintsov_2020}
S.~Odintsov and V.K.~Oikonomou, \emph{Geometric inflation and dark energy with
  axion f(r) gravity},
  \href{https://doi.org/10.1103/physrevd.101.044009}{\emph{Physical Review D}
  {\bfseries 101} (2020) }.

\bibitem{Oikonomou_2022}
V.K.~Oikonomou, \emph{Kinetic axion f(r) gravity inflation},
  \href{https://doi.org/10.1103/physrevd.106.044041}{\emph{Physical Review D}
  {\bfseries 106} (2022) }.

\bibitem{Visinelli:2017ooc}
L.~Visinelli, S.~Baum, J.~Redondo, K.~Freese and F.~Wilczek, \emph{{Dilute and
  dense axion stars}},
  \href{https://doi.org/10.1016/j.physletb.2017.12.010}{\emph{Phys. Lett. B}
  {\bfseries 777} (2018) 64}
  [\href{https://arxiv.org/abs/1710.08910}{{\ttfamily 1710.08910}}].

\bibitem{Zhang:2018slz}
H.~Zhang, \emph{Axion stars},
  \href{https://doi.org/10.3390/sym12010025}{\emph{Symmetry} {\bfseries 12}
  (2020) }.

\bibitem{Guth_2015}
A.H.~Guth, M.P.~Hertzberg and C.~Prescod-Weinstein, \emph{Do dark matter axions
  form a condensate with long-range correlation?},
  \href{https://doi.org/10.1103/physrevd.92.103513}{\emph{Physical Review D}
  {\bfseries 92} (2015) }.

\bibitem{Eby:2015hyx}
J.~Eby, P.~Suranyi and L.C.R.~Wijewardhana, \emph{{The Lifetime of Axion
  Stars}}, \href{https://doi.org/10.1142/S0217732316500905}{\emph{Mod. Phys.
  Lett. A} {\bfseries 31} (2016) 1650090}
  [\href{https://arxiv.org/abs/1512.01709}{{\ttfamily 1512.01709}}].

\bibitem{Zhang:2020bec}
H.-Y.~Zhang, M.A.~Amin, E.J.~Copeland, P.M.~Saffin and K.D.~Lozanov,
  \emph{{Classical Decay Rates of Oscillons}},
  \href{https://doi.org/10.1088/1475-7516/2020/07/055}{\emph{JCAP} {\bfseries
  07} (2020) 055} [\href{https://arxiv.org/abs/2004.01202}{{\ttfamily
  2004.01202}}].

\bibitem{Zhang_2021}
H.-Y.~Zhang, \emph{Gravitational effects on oscillon lifetimes},
  \href{https://doi.org/10.1088/1475-7516/2021/03/102}{\emph{Journal of
  Cosmology and Astroparticle Physics} {\bfseries 2021} (2021) 102}.

\bibitem{PhysRevLett.117.121801}
E.~Braaten, A.~Mohapatra and H.~Zhang, \emph{Dense axion stars},
  \href{https://doi.org/10.1103/PhysRevLett.117.121801}{\emph{Phys. Rev. Lett.}
  {\bfseries 117} (2016) 121801}.

\bibitem{Sen:2021mhf}
S.~Sen and L.~Sivertsen, \emph{{Electromagnetic radiation from axion
  condensates in a time dependent magnetic field}},
  \href{https://doi.org/10.1007/JHEP05(2022)192}{\emph{JHEP} {\bfseries 05}
  (2022) 192} [\href{https://arxiv.org/abs/2111.08728}{{\ttfamily
  2111.08728}}].

\bibitem{Arvanitaki_2010}
A.~Arvanitaki, S.~Dimopoulos, S.~Dubovsky, N.~Kaloper and J.~March-Russell,
  \emph{String axiverse},
  \href{https://doi.org/10.1103/physrevd.81.123530}{\emph{Physical Review D}
  {\bfseries 81} (2010) }.

\bibitem{Amin2021}
M.A.~Amin, A.J.~Long, Z.-G.~Mou and P.M.~Saffin, \emph{Dipole radiation and
  beyond from axion stars in electromagnetic fields},
  \href{https://doi.org/10.1007/JHEP06(2021)182}{\emph{Journal of High Energy
  Physics} {\bfseries 2021} (2021) 182}.

\bibitem{Amin:2021tnq}
M.A.~Amin, A.J.~Long, Z.-G.~Mou and P.~Saffin, \emph{{Dipole radiation and
  beyond from axion stars in electromagnetic fields}},
  \href{https://doi.org/10.1007/JHEP06(2021)182}{\emph{JHEP} {\bfseries 06}
  (2021) 182} [\href{https://arxiv.org/abs/2103.12082}{{\ttfamily
  2103.12082}}].

\bibitem{Schiappacasse_2018}
E.D.~Schiappacasse and M.P.~Hertzberg, \emph{Analysis of dark matter axion
  clumps with spherical symmetry},
  \href{https://doi.org/10.1088/1475-7516/2018/01/037}{\emph{Journal of
  Cosmology and Astroparticle Physics} {\bfseries 2018} (2018) 037}.

\bibitem{Piette_1998}
B.~Piette and W.J.~Zakrzewski, \emph{Metastable stationary solutions of the
  radial d-dimensional sine-gordon model},
  \href{https://doi.org/10.1088/0951-7715/11/4/020}{\emph{Nonlinearity}
  {\bfseries 11} (1998) 1103}.

\bibitem{alma990016848140102756}
W.H.~Press, S.A.~Teukolsky, W.T.~Vetterling and B.P.~Flannery, \emph{Numerical
  Recipes 3rd Edition: The Art of Scientific Computing}, Cambridge University
  Press, 3~ed. (2007).

\bibitem{PhysRevD.100.063002}
J.~Eby, M.~Leembruggen, L.~Street, P.~Suranyi and L.C.R.~Wijewardhana,
  \emph{Global view of qcd axion stars},
  \href{https://doi.org/10.1103/PhysRevD.100.063002}{\emph{Phys. Rev. D}
  {\bfseries 100} (2019) 063002}.

\bibitem{Grandcl_ment_2011}
P.~Grandcl{\'{e} }ment, G.~Fodor and P.~Forg{\'{a}}cs, \emph{Numerical
  simulation of oscillatons: Extracting the radiating tail},
  \href{https://doi.org/10.1103/physrevd.84.065037}{\emph{Physical Review D}
  {\bfseries 84} (2011) }.

\bibitem{PhysRevLett.66.1659}
E.~Seidel and W.-M.~Suen, \emph{Oscillating soliton stars},
  \href{https://doi.org/10.1103/PhysRevLett.66.1659}{\emph{Phys. Rev. Lett.}
  {\bfseries 66} (1991) 1659}.

\bibitem{Ure_a_L_pez_2002}
L.A.U.~a~L~pez, T.~Matos and R.~Becerril, \emph{Inside oscillatons},
  \href{https://doi.org/10.1088/0264-9381/19/23/320}{\emph{Classical and
  Quantum Gravity} {\bfseries 19} (2002) 6259}.

\bibitem{Alcubierre_2003}
M.~Alcubierre, R.~Becerril, F.S.G.~n, T.~Matos, D.~o~N~ez and L.A.U.~a~L~pez,
  \emph{Numerical studies of $\phi^2$ oscillatons},
  \href{https://doi.org/10.1088/0264-9381/20/13/332}{\emph{Classical and
  Quantum Gravity} {\bfseries 20} (2003) 2883}.

\bibitem{Kolokolov1973}
A.A.~Kolokolov, \emph{Stability of the dominant mode of the nonlinear wave
  equation in a cubic medium},
  \href{https://doi.org/10.1007/BF00850963}{\emph{Journal of Applied Mechanics
  and Technical Physics} {\bfseries 14} (1973) 426}.

\bibitem{LEE1992251}
T.~Lee and Y.~Pang, \emph{Nontopological solitons},
  \href{https://doi.org/https://doi.org/10.1016/0370-1573(92)90064-7}{\emph{Physics
  Reports} {\bfseries 221} (1992) 251}.

\bibitem{Nugaev2020}
E.Y.~Nugaev and A.V.~Shkerin, \emph{Review of nontopological solitons in
  theories with u(1)-symmetry},
  \href{https://doi.org/10.1134/S1063776120020077}{\emph{Journal of
  Experimental and Theoretical Physics} {\bfseries 130} (2020) 301}.

\bibitem{Sen:2018cjt}
S.~Sen, \emph{{Plasma effects on lasing of a uniform ultralight axion
  condensate}}, \href{https://doi.org/10.1103/PhysRevD.98.103012}{\emph{Phys.
  Rev. D} {\bfseries 98} (2018) 103012}
  [\href{https://arxiv.org/abs/1805.06471}{{\ttfamily 1805.06471}}].

\end{thebibliography}\endgroup
\end{document}